\documentclass[12pt]{article}
\usepackage{graphicx}
\begin{document}
\title{New physics in frustrated magnets: Spin ices, monopoles, etc.}
\vspace{1.5em}
\author{A.A. Zvyagin$^{(1,2)}$ \\ 
$^{(1)}$ Max Planck Institut f\"ur Physik komplexer Systeme, \\
N\"othnitzer Str. 38, D-01187 Dresden, Germany \\ 
$^{(2)}$ B.I.~Verkin Institute for Low Temperature Physics and \\
Engineering, Ukrainian National Academy of Sciences, \\
47 Lenin Avenue, Kharkov, 61103, Ukraine }
\date{\today}
\maketitle
\begin{abstract} 
During recent years the interest to frustrated magnets has grown considerably. 
Such systems reveal very peculiar properties which distinguish them 
from standard paramagnets, magnetically ordered regular systems (like ferro-, 
ferri-, and antiferromagnets), or spin glasses. In particular great 
amount of attention has been devoted to the so-called spin ices, in which 
magnetic frustration together with the large value of the single-ion magnetic 
anisotropy of a special kind, yield peculiar behavior. One of the most 
exciting features of spin ices is related to low-energy emergent excitations, 
which, from many viewpoints can be considered as analogies of Dirac's 
monopoles. In this article we review the main achievements of theory and 
experiment in this field of physics.   
\end{abstract}


\newpage

\section{Introduction}

Magnetic materials are among the oldest systems that have been studied by 
physicists \cite{Mat}. The interest to magnetic systems is connected with 
their special properties. Also the application of magnetic materials in modern 
technology put the study of such systems to the one of the main aims of modern 
condensed matter physics. On the other hand, theoretical models, which 
originally were developed to describe magnetic properties of matter, like the 
famous Ising model, are often used in other fields of theoretical and 
experimental physics. The opposite is also true: Many approaches of modern 
physics are successfully used in the theory and experiment of magnetism. 

One of the advantages of the theory of magnetism, is the well-developed during 
years conceptual approach there \cite{Zb}. For example, at the classical 
level, Maxwell's electrodynamics has successfully described the main features 
of the response of magnetic materials to the external electric and magnetic 
fields. On the other hand, the quantum nature of magnetism manifests itself, 
e.g., in properties of non-interacting with each other magnetic ions in 
paramagnets. Schottky anomalies in the behavior of magnetic contribution to 
the specific heat are the prime example of the quantum nature of 
(para)magnetic ions due to the crystalline electric field of ligands. The 
theory of such paramagnets is well-developed \cite{Ball,vV,AB1,Stev}. In 
general, we know how interactions between magnetic ions change their 
properties (starting with present in any magnetic system magnetic 
dipole-dipole interactions, short-range exchange interactions 
\cite{Heis,Dir1}, which mostly define magnetic ordering, and long-range 
magnetic interactions in metals \cite{RK,Kas,Yos}, which 
can often be the reason for inhomogeneity in magnetic structures). 

\begin{figure}
\begin{center}
\vspace{-15pt}
\includegraphics[scale=0.1]{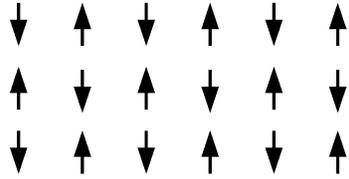}
\end{center}
\vspace{-15pt}
\caption{The ground state for the spin-1/2 antiferromagnetic Ising model on 
the square lattice.}   
\label{afIs}
\end{figure}

For standard many-body magnets we know how to take into account interactions. 
As a rule we successfully use the mean-field-like theory \cite{Smart}, or at 
low temperatures, the spin wave approximation \cite{spinwave,HolPrim}. Such 
theories can be used without principal difficulties for systems, in which we 
can well determine the ground state, i.e., the optimal state with the minimal 
energy, like in ferromagnets or two-sublattice antiferromagnets 
\cite{Neel,Neel1}. However, if the situation with inter-ionic magnetic 
interactions becomes more complicated than in standard bipartite magnetic 
systems, where the nearest neighbor antiferromagnetic interactions can be 
satisfied for each pair of magnetic particles, like in the square lattice 
Ising antiferromagnet, see Fig.~\ref{afIs}, standard mean field and spin wave 
methods cannot be applied successfully, and we need different approaches.     

In bipartite lattices we can divide the total system in two subsystems so 
that particles, belonging to the first subsystem, are nearest neighbors to 
particles, belonging to the other subsystem. If the interaction is between 
only nearest neighbors, pair antiferromagnetic bonds have minimal energies 
and the global optimal state of the system can be realized by minimizing 
coupling energies for each pair. However, there exist many lattices, which 
we cannot divide into two sublattices. For such systems we have a problem 
with the use of mean-field-like approximation or spin wave theory: The optimal 
state with the minimal energy is either not determined there, or there are 
many such states (too many, their number is of order of number of magnetic 
particles in the system), so that we cannot realize the knowledge of the 
ground state. The simple example is the triangular two-dimensional lattice. The 
elementary cell of the triangular lattice is a triangle. 

\begin{figure}
\begin{center}
\vspace{-15pt}
\includegraphics[scale=0.3]{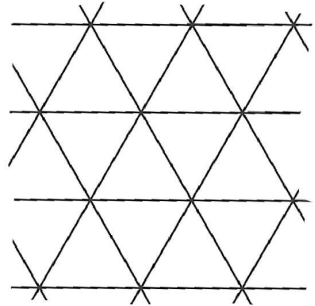}
\end{center}
\vspace{-15pt}
\caption{The example of non-bipartite two-dimensional lattice: 
triangular lattice.}   
\label{triangle}
\end{figure}

Another example of such a lattice is the so-called Kagome lattice, known due to 
traditional Japanese bamboo baskets. It is composed of the arrangement of 
interlaced triangles, which are organized so that each point where two laths 
cross has four neighboring pattern of a tri-hexagonal tiling. The elementary 
cell of the Kagome lattice is the star of David.

\begin{figure}
\begin{center}
\vspace{-15pt}
\includegraphics[scale=0.1]{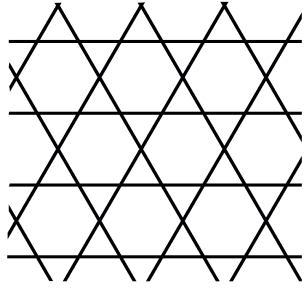}
\end{center}
\vspace{-15pt}
\caption{The example of non-bipartite two-dimensional lattice: 
Kagome lattice.}   
\label{Kagome}
\end{figure}

Notice that the crossing points of the Kagome lattice do not form a 
mathematical lattice, unlike the triangular lattice \cite{Con}. It has the 
symmetry p6m (or p3m1), like the triangular lattice. We can see that 
antiferromagnetic nearest-neighboring couplings cannot minimize the total 
energy of such a system. Hence, the standard approach, which brought so much 
success in studies of bipartite (antiferro)magnetic systems, fails for such 
lattices. Such magnetic systems are known nowadays as magnetically frustrated 
ones. In many magnetically frustrated systems magnetic ions do not develop 
long-range magnetic ordering for the reasons, which will be explained below. 
In that sense frustrated magnetic systems belong to the class of 
``spin-liquids'' \cite{And1,Bal1}. The quantum spin liquid state is 
disordered, like in liquids, comparing to magnetically long-range ordered 
states. However, unlike other disordered states, a spin liquid state can be 
preserved down to very low temperatures (comparing to the values of spin-spin 
interactions). The interest to magnetically frustrated systems is caused not 
only by their interesting physical properties; such materials are perspective 
from the point of view of their use as data storage and memory, or as possible 
realization of topological quantum computation. 
 
\section{Frustration}

We call the system as frustrated if it cannot minimize its total energy (the 
macroscopic state) by minimizing the interaction between each pair involved 
into the interaction, i.e., to perform such a minimization pair by pair 
\cite{Toul,Vill}. On the other hand, it is often used to call the system 
frustrated if its ground state is highly degenerate, and the level of 
degeneracy is of order of the number of particles in the system. Magnetic 
systems are the most known example of the manifestation of frustration. \
However, naturally, the phenomenon of frustration is not limited to magnetic 
systems. For example, among frustrated systems we can count liquid and 
molecular crystals (like solid N$_2$), arrays of Josephson junctions, as well 
as the so-called ``nuclear pasta state'' of spatially modulated nuclear 
density inside stars (caused by the competition between Coulomb interactions 
and short-range nuclear couplings). 

\begin{figure}
\begin{center}
\vspace{-15pt}
\includegraphics[scale=0.5]{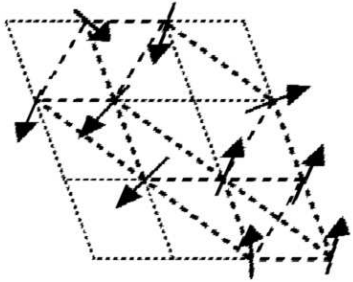}
\end{center}
\vspace{-15pt}
\caption{Illustration of a spin glass state: Spins are randomly distributed 
on a regular lattice.}   
\label{spinglass}
\end{figure}

It is usual to distinguish between random and geometrical frustrations. Let 
us, first, discuss in short the former, because our review is mainly devoted 
to the latter. The random frustration, in turn, can be divided into dynamical 
and quenched one, by the origin. The characteristic feature of the dynamical 
(annealed) random frustration is related to multiple length scales, which are 
developed in time for spatially inhomogeneous systems with competing 
interactions. If such dynamical processes are frozen out, the randomness, and, 
in turn, frustration, is quenched. To remind, in statistical physics we usually 
call some parameters as quenched when they are random variables which do not 
evolve in time. Quenched frustration appears in systems, in which frozen 
degrees of freedom are not homogeneous, e.g., they cannot be periodically 
translated. Such a phenomenon can be observed in many metallic alloys with 
magnetic ingredients, which interact with each other via the long-range 
sign-changing Rudermann-Kittel-Kasuya-Yoshida (RKKY) coupling 
\cite{RK,Kas,Yos}. The main example of the manifestation of the random 
frustration in magnetic systems is a spin glass 
\cite{sg,EdAnd,SherKirk,MezParVir,FishHer,Myd}, i.e., an ordered magnet with 
stochastic positions of spins with competing possible ferromagnetic and 
antiferromagnetic interactions between them, see the example in 
Fig.~\ref{spinglass}. 

The spontaneous magnetization of spin glasses is zero, however the magnetic 
ordering exists in the form of long-ranged spin-spin correlations. 
 
\begin{figure}
\begin{center}
\vspace{-15pt}
\includegraphics[scale=0.6]{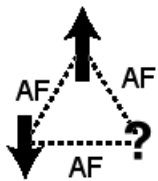}
\end{center}
\vspace{-15pt}
\caption{Elementary cell of the antiferromagnetic spin-1/2 Ising model on the 
two-dimensional equilateral triangular lattice. It is impossible to satisfy 
all three antiferromagnetic bonds simultaneously.}   
\label{tri}
\end{figure}

In this review we will mostly deal with the geometric frustration. Here 
particles sit on the sites of regular lattices, unlike the situation with 
random frustration. However, local pair particle-particle interactions are in a 
conflict with each other: Each bond favors its own spatial correlation. Then it 
is impossible to satisfy all local interactions. The most known example 
is related to Ising spins 1/2 (which can be directed only up and down); they 
interact antiferromagnetically only with nearest neighbors on a 
two-dimensional equilateral triangular lattice. Clearly, antiferromagnetic 
bonds tend each neighboring spins to be antiparallel to each other, but it is 
impossible to realize, hence frustration. An example of the elementary cell of 
such a system is presented in Fig.~\ref{tri}.  

Geometrical frustration is possible not only if spins are collinear, but for 
spins arranged non-collinearly. Geometrically frustrated systems often manifest 
a residual entropy. The residual entropy, by definition, is the amount of 
entropy present even if the system is cooled arbitrary close to zero 
temperature. It exists for systems, in which many different microscopic states 
can persist when cooled to zero temperature, e.g., if the system has many 
different ground states with the same energy: degenerate ground states. Such a 
situation can also exist if such states have slightly different energies, but 
the system is prevented from settling in the ``real'' ground state with the 
lowest energy. The latter can be realized, e.g., if the system is very swiftly 
cooled. The most known example for systems possessing residual entropy is any 
amorphous system, like a glass. There the reason for residual entropy is 
caused in a great number of different ways of realization of microscopic 
structures in a macroscopic system. The interesting property of geometrically 
frustrated magnetic systems, like spin ice (see below) is that the level of 
residual entropy can be controlled by the application of an external magnetic 
field. This property of geometrically frustrated magnetic systems can be used 
for creation of refrigeration systems. In fact, geometrically frustrated 
magnetic systems had been studied earlier than the term ``frustration'' has 
been used \cite{Hou,Wan,YafKit}. For the review on frustrated spin systems, 
consult, e.g., the interesting books \cite{Diep,Lacr}. Perhaps, it is 
worthwhile to discuss here the convenient measure of the level of frustration 
in geometrically or randomly frustrated magnetic systems. So-called frustration 
index $f$ has been proposed \cite{Ram}. It is determined as $f = 
|\theta_{\rm CW}|/T_c$, where $\theta_{\rm CW}$ is the Curie-Weiss temperature, 
which can be extracted from the temperature behavior of the inverse magnetic 
susceptibility, and $T_c$ is the (critical) temperature at which the magnetic 
system possesses the long range order (say, the N\'eel temperature for 
antiferromagnets, or freezing temperature for spin glasses). Clearly, for 
magnetically disordered frustrated spin systems we would have $f \to \infty$. 
However, in the most of real magnetic systems spin-spin interactions (for 
example, magnetic dipole-dipole interactions, which are present in any magnetic 
system) should develop magnetic ordering, though at very low temperatures. 
Geometrically frustrated magnetic systems have been reviewed in 
\cite{Ram1,Gaul,ShifRam,Geed,MoeRam,CMS1}.

Probably, the oldest example of the geometrically frustrated system is the 
usual water ice. 

\section{Water Ice}

It is well known that the molecule of water consists of two hydrogen atoms 
connected with the help of a covalent bond to the oxygen atom. Water ice is the 
frozen water, i.e., it is the water in the solid state. Depending on the 
external temperature and pressure water molecules in (water) ice can be 
organized in different forms. At ambient pressure the water ice can exist in 
three common forms: the ice I$_{\rm h}$, or the hexagonal ice, which possesses 
the hexagonal symmetry, the most common phase of the water ice; the ice 
I$_{\rm c}$, or the cubic ice, in which the cubic symmetry persists, and the 
ice XI with the orthorhombic symmetry (the space group Cmc2$_1$) \cite{strXI}. 
The ice I$_{\rm c}$ or sphalerite, is the metastable phase existing, as a rule, 
between 130~K and 220~K, in which oxygen atoms organize a cubic diamond 
structure \cite{Mur}. The ice XI is the proton (hydrogen)-ordered 
low-temperature (below 72~K) form of the hexagonal ice. It contains eight 
water molecules per unit cell. The internal energy of the ice XI is about 0.17 
times lower than the one for the hexagonal ice I$_{\rm h}$. It is a 
ferroelectric, see, e.g., \cite{fer}. 

The hexagonal ice, also known as ice one, or wurzite, is the water ice, which 
properties permitted to give the name to spin ices. It is stable down to 
approximately 73~K. Its symmetry is hexagonal with nearly tetrahedral bonding 
angles ($\arccos(-1/3) \approx 109.5^{\rm o}$) of the crystal structure. The 
latter consists of crinkled (alternating in the ABAB pattern) planes composed 
of tessellating hexagonal rings (a repetition of rings without gaps and 
overlaps, like in Escher's pictures). B planes are reflections of A planes 
along the same axes as the planes themselves. Oxygen sits in each vertex, and 
edges of rings are formed by hydrogen bonds \cite{hbond,hbond1}. 

\begin{figure}
\begin{center}
\vspace{-15pt}
\includegraphics[scale=0.7]{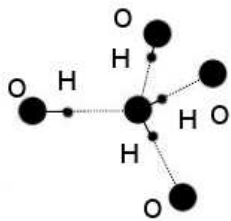}
\end{center}
\vspace{-15pt}
\caption{Configuration of oxygens and hydrogens in the hexagonal water ice 
(oxygen: large spheres, hydrogen: small spheres). Each oxygen-oxygen bond has 
two steady-state positions of hydrogen. The configuration satisfies 
Bernal-Fowler ice rules: One hydrogen per oxygen-oxygen bond, and each oxygen 
neighbors two close hydrogens and two sitting far hydrogens.}   
\label{Ihice}
\end{figure}

Two molecules of water can form a hydrogen bond between them. In a liquid 
water more bonds are possible because oxygen of a single water molecule has 
two lone pairs of electrons, each can form a hydrogen bond with another 
molecule with the angle between hydrogen atoms 104.45$^{\rm o}$ and with the 
distance from hydrogen to oxygen being 95.84~pm. The side with oxygen atom in 
the water molecule has a partial negative charge due to higher 
electronegativity of the oxygen comparing to hydrogen. It means that the 
hydrogen side is partially positive, i.e., the 
water molecule is a dipole. The charge difference yields attraction between 
water molecules, which contributes to the hydrogen bonding. Every water 
molecule has hydrogen bonds with up to four other water molecules, because it 
can accept two and donate two hydrogens, see Fig.~\ref{Ihice}. The hydrogen 
bonding energy of the water molecule is relatively strong (it is weak, though, 
comparing to covalent bonds within the water molecule). Hydrogen bonds with 
almost tetrahedral bonding angles of the water molecule, cf. Fig.~\ref{Ihice},
help to organize an open hexagonal lattice of the hexagonal ice. The distance 
between oxygen atoms along each bond is about 275~pm, which is much larger 
than the distance between oxygen and hydrogen in the water molecule. Large 
hexagonal rings leave almost enough room for another water molecule to exist 
inside, which yields the density of ice being lower than of the water. 

Hydrogen atoms (in fact, almost protons) sit very close along hydrogen bonds 
in the crystal lattice of the hexagonal ice (Fig.~\ref{Ihice}), i.e., each 
water molecule is preserved there. It implies that in the hexagonal ice each 
oxygen has two adjacent hydrogens at about 101~pm (along the 275~pm hydrogen 
bond), i.e., not in the middle of the distance between two oxygens. Basically, 
two equivalent hydrogen positions exist in each oxygen-oxygen bond. Four-fold 
oxygen coordination yields one hydrogen per such a bond. In the structure of 
the hexagonal water ice that way is determined by the Bernal-Fowler ice rules 
\cite{BF}. The first one is related to one hydrogen per oxygen-oxygen bond in 
average in the ice crystal. The second ice rule states that for each oxygen 
two hydrogens have to be close, and two protons sit far from the oxygen. It 
turns out that the second ice rule frustrates the low-energy problem of the 
water ice caused by the stability of water molecules in it. As a result, the 
crystal structure contains the residual (zero temperature) entropy inherent to 
the lattice. In other words, the hexagonal ice is expected to have the 
intrinsic randomness even if it was possible to cool it to zero temperature. 
In ideal situation the hexagonal (water) ice can never be completely frozen, 
seemingly violating the third law of thermodynamics! Such an entropy 
in the hexagonal water ice is defined by the number of possible configurations 
of hydrogen positions which can be formed (the requirement of two hydrogens to 
be related to each oxygen in the closest proximity with each hydrogen bond, 
which join two oxygen atoms having only one hydrogen, holds). The residual 
entropy of the hexagonal ice is ${\cal S}_0 \approx 3.5$~J~mol$^{-1}$~K$^{-1}$. 
That value has been measured in the set of experiments devoted to the 
investigation of the specific heat of the hexagonal water ice \cite{Gia,Gia1}.

The structure of the hexagonal ice has been pioneered by Linus Pauling [the 
only person who was awarded by two unshared Nobel Prizes: Chemistry (1954) 
and Peace (1962) prizes] in 1935 \cite{Paul}. He has noticed that the number 
of configurations with two hydrogen being close to the oxygen, and two 
hydrogens being far from it grows exponentially with the system size. It 
implies the extensive character of the residual entropy of the hexagonal water 
ice. Pauling has estimated the value of the residual entropy in the hexagonal 
water ice. One mole of ice contain $N$ oxygens, and therefore $2N$ 
oxygen-oxygen bonds. Each such a bond can have two possible positions for a 
hydrogen, which implies $2^{2N}$ possible hydrogen positions for the total 
crystal. Only six configurations are energetically favorable out of 16 
possible ones for each oxygen. The upper limit for a number of ground state 
configurations, $M$, can be, therefore, estimated as $2^N (6/16)^N = (3/2)^N$. 
Corresponding entropy can be calculated as ${\cal S}_0 = N k_{\rm B}\ln (3/2)$, 
which gives 3.37~J~mol$^{-1}$~K$^{-1} \approx 0.323 k_{\rm B}$. That value 
agrees very well with the experimentally measured \cite{Gia,Gia1}. Despite 
calculations performed by Pauling missed the global constraint of the number 
of hydrogens and local constraints caused by closed loops in the lattice of 
the hexagonal water ice, its accuracy is of order of 1-2 \% \cite{Nag}. It has 
been calculated numerically for the two- and three-dimensional ice model (see 
below). 

\subsection{Ice Models}

\begin{figure}
\begin{center}
\vspace{-15pt}
\includegraphics[scale=0.19]{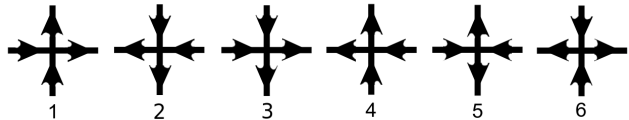}
\end{center}
\vspace{-15pt}
\caption{Six possible configurations of the square ice model, possessing ice 
rules: In each vertex two arrows point inwards and two arrows point 
outwards.}   
\label{6vert}
\end{figure}

Ice-type models, i.e., the ones, which possess ice rules, are often studied 
in statistical mechanics: They are the particular case of vortex models, 
namely, the six-vortex models. Any ice model is defined on a lattice with 
the coordination number 4, i.e., each vertex is connected to four nearest 
neighbors by an edge. Each bond is represented by an arrow, so that the 
number of arrows pointing to the vortex is two (as well as the number of 
arrows pointing outwards), which constitutes the ice rule in the vertex 
model. So far, mostly two- and three-dimensional ice vertex models has been 
studied. For instance, for the square ice model six configurations are valid.
The energy of the state $E$ is given by $E=\sum_{i=1}^6\varepsilon_in_i$, 
where $n_i$ is the number of vertices with $i$-th configuration (of six 
possibilities), and $\varepsilon_i$ being the energy associated with the 
vertex configuration $i$. Fig.~\ref{6vert} shows six possible configurations 
of the six-vertex square model, which satisfy the ice rule. 

The six-vertex model on a square lattice can model a ferroelectric \cite{Rys}, 
where $\varepsilon_{1,2,3,4} >0$ and $\varepsilon_{5,6}=0$. If there is no 
external field, the condition $\varepsilon_1=\varepsilon_2$, 
$\varepsilon_3=\varepsilon_4$ and $\varepsilon_5=\varepsilon_6$ holds. The 
six-vertex model on a square lattice has been exactly solved by 
E.~Lieb \cite{Lieb,Lieb1,Lieb2}. He has found the residual 
entropy there ${\cal S}_0 = (3/2)Nk_{\rm B}\ln (4/3)$, and the value 
$(4/3)^{3/2} \approx 1.5396$ is known as the Lieb's square ice constant. Later 
Lieb's solution has been generalized for the cases with \cite{Suth} and 
without external field \cite{Y}. Naturally, one can consider more realistic 
models, than the two-dimensional six-vertex model on the square lattice. 
However, for the three-dimensional ice-type model the exact solution has been 
obtained \cite{N} only for the special temperature interval, where the model 
is called to be ``frozen'', which means that in the thermodynamic limit the 
energy and entropy per vortex are zero in such a range of $T$. Ice-type 
vertex models in statistical mechanics are generalized for the eight-vertex 
model, which also possesses exact solution \cite{Bax}.

\section{Spinels: Cation ordering and antiferomagnetic models} 

\begin{figure}
\begin{center}
\vspace{-15pt}
\includegraphics[scale=0.5]{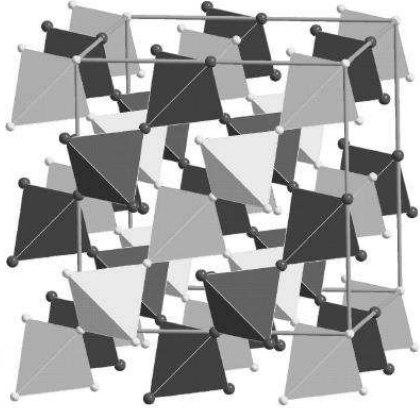}
\end{center}
\vspace{-15pt}
\caption{A- (light tetrahedra) and B-cites (dark tetrahedra) of 
A$_2$B$_2$O$_7$ in the vertices of corner-sharing tetrahedra form the 
pyrochlore lattice. From J.S.~Gardner {\em et al.}, Physical Review B 
{\bf 70}, 180404(R) (2004). http://link.aps.org/abstract/PRB/v70/p180404   
Copyright 2004 by the American Physical Society.}   
\label{pyrochlore}
\end{figure}

The similarity of the water ice problem to the ordering of cations in the 
so-called inverse spinel material has been pointed out by E.J.W.~Verwey 
\cite{Ver,VerHaa}, and then has been discussed in detail by P.W.~Anderson 
\cite{And}. Spinels (called due to the natural mineral spinel MgAl$_2$O$_4$) 
are the class of materials with the general chemical formula AB$_2$O$_4$ with 
the cubic crystal system. A and B cations occupy octahedral and tetrahedral 
sites of the lattice, and can be divalent, trivalent or quadrivalent. In 
inverse spinels two kinds of cations on the B-sites of the spinel lattice are 
situated so that the total numbers of cations of each kind are equal. The 
B-sites of the spinel lattice form the so-called pyrochlore lattice. The latter 
[called due to pyrochlore, the natural mineral with the chemical formula 
(Na,Ca)$_2$Nb$_2$O$_6$(OH,F)] has Fd${\bar 3}$m space group, often is related to
systems with the chemical formulas A$_2$B$_2$O$_6$ or A$_2$B$_2$O$_7$. The 
pyrochlore lattice is organized of corner-sharing tetrahedra, which are 
alternating ``upward'' and ``downward'', see Fig.~\ref{pyrochlore}. 

The minimum energy is related to the 
case, in which the number of pairs consisting of two different kinds of 
cations is maximal. Such a condition is satisfied if each elementary 
tetrahedron of the B-lattice of the inverse spinel material has two cations of 
one kind and two cations of the other kind, so called tetrahedron rule, 
analogous to the ice rule for the hexagonal water ice. Notice that in the 
spinel lattice centers of tetrahedra are situated on the same lattice as 
oxygens in the cubic I$_{\rm c}$ water ice. It implies that cation ordering in 
this problem could have residual entropy, like in the Pauling water ice. 

Among spinels with different cations at B-sites we can distinguish the 
situation, in which the valency of ions is different from integer, like in 
the first non-rare-earth based heavy-fermion system LiV$_2$O$_4$ see, e.g., 
\cite{Li,KonJonMil,Chmai,Ey,KriLoi,FYZG,LeeLV,YamU,MatsUU,ShiTak}, or in, 
probably, the oldest known magnetic material, magnetite, Fe$_3$O$_4$ see, 
e.g., \cite{Mat,Ver,VerHaa}. There the formal valence of V ion is 3.5, and the 
one of Fe on B sites is 2.5 (the valence of Fe ions on A sites is 3), i.e., 
they have to exist in equal combinations of V$^{3+}$ and V$^{4+}$, 
or Fe$^{2+}$ and Fe$^{3+}$. Notice, however, that recent studies contradict the 
direct application of the ice (tetrahedron) rule to LiV$_2$O$_4$ and 
Fe$_3$O$_4$  
\cite{Li,KonJonMil,Chmai,Ey,KriLoi,FYZG,LeeLV,YamU,MatsUU,ShiTak,mag,Sam,Chib,
MiyShi,WrAttRad}. For the recent reviews on spinel materials 
consult, e.g., \cite{revspin,Lac,LeeTak,TakU}. 

Similar situation appears if we consider the spin Ising antiferromagnetic 
model on the pyrochlore lattice. Here spin up and spin down correspond to two 
kinds of cations in the above mentioned spinel situation, or to ``close 
hydrogen'' or ``far hydrogen''for the water ice. However, there is no 
realization of such an Ising model on the pyrochlore lattice. Why is it so? 
The pyrochlore lattice has the cubic symmetry. Hence, there is no reason for 
the unique direction of Ising spins in such a system. On the other hand, the 
situation with antiferromagnetic Heisenberg spins on a pyrochlore lattice is 
realistic. It was J.~Villain, who pointed out that the classical Heisenberg 
antiferromagnet cannot be magnetically ordered on the pyrochlore lattice 
\cite{Vil} due to the geometrical frustration down to zero temperatures. He 
called such system as a {\em collective paramagnet} to stress the absence of 
ordering, on the one hand, and collective nature of magnetic properties, on the 
other hand. 

\section{Spin ice}

\begin{figure}
\begin{center}
\vspace{-15pt}
\includegraphics[scale=0.6]{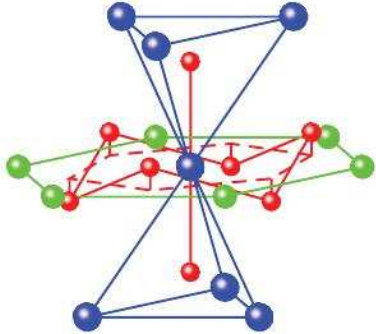}
\end{center}
\vspace{-15pt}
\caption{(Color online) The environment of the A-site of pyrochlore oxides. 
Large (blue) spheres denote rare-earth ions, medium (green) spheres denote 
nonmagnetic metal ions, and small (red) spheres are oxygens. Six oxygens 
(O) lie in plane, and two (O') are situated on the line, perpendicular to 
basal plane. From A.~Yaouanc {\em et. al.}, Physical Review B {\bf 84}, 172408 
(2011). http://link.aps.org/abstract/PRB/v84/p172408 
Copyright 2011 by the American Physical Society.}   
\label{Asiteen}
\end{figure}

Magnetic systems, in which magnetic ions reside on lattices of corner-sharing 
tetrahedra (pyrochlore lattice), belong to the most known examples of 
magnetic systems with geometrical frustration. Among them, maybe the most 
interesting properties are revealed by the cubic pyrochlore oxides of the 
family A$_2$B$_2$O$_7$ with magnetic A ions and nonmagnetic B ones 
\cite{Sub,Gree,G,GGG}. Such systems can be metallic, or insulating. The space 
group for that systems is Fd${\bar 3}$m. It is usual to use another chemical 
formula, namely A$_2$B$_2$O$_6$O' to emphasize the difference in positions of 
oxygen ions. Here A ion is placed in 16d position in the Wyckoff 
classification [minimal coordinates are (1/2,1/2,1/2)], B ion is in 
16c placed at the origin [i.e., minimal coordinates are (000)], O is in 48f 
(x,1/8,1/8) and O' is in 8b (3/8,3/8,3/8). Here the parameter x is of range 
0.32-0.345. All six B-O bonds have equal lengths, and O-B-O angles have almost 
ideal octahedral values of 90$^{\rm o}$, i.e., oxygens surround B-ion at 
vertices of the perfect octahedron. As for A-ion, oxygen ions form the perfect 
cube, but with strong distortions. In fact, the surrounding of the A-site can 
be considered as six-membered ring of O with two O' atoms, which form a stick, 
oriented perpendicular to the ring, see Fig.~\ref{Asiteen}. 

This is why, A-ions have a large axial symmetry, with the axes parallel to 
[111] directions (diagonals of the cube). Basically, such a axial symmetry 
produces a large crystalline electric field at A site, which is the origin of 
the Ising-like properties of magnetic ions situated at the A position.  

Probably, the most interesting representatives of pyrochlore oxides are the 
ones with A being trivalent rare-earth ions, like Gd, Tb, Dy, Ho, Yb, etc., 
or Y, and with a tetravalent ion in B-position, as Ti, Sn, Mo, Mn, etc. Both A 
and B ions can be magnetic or non-magnetic.   

We will concentrate now on insulating titanates of Dy and Ho \cite{Ginrev}
(Dy$_2$Ti$_2$O$_7$, and Ho$_2$Ti$_2$O$_7$), and on similar compounds, 
stannates, with the replacement of non-magnetic Ti$^{4+}$ by the non-magnetic 
Sn$^{4+}$. In this case only A-sublattice (which is the FCC lattice 
of corner-sharing tetrahedra, directed up and down) is responsible for magnetic 
properties. The primitive basic cell has four rare-earth ions sitting at the 
vertices of each tetrahedron. However, the conventional cubic unit cell of 
pyrochlore oxides has the size $a \sim 10$~\AA with 16 rare-earth ions, i.e., 
it consists of four primitive tetrahedron cells directed ``up'' and ``down''. 
The distance between nearest neighboring rare-earth ions is 
$x = \sqrt{2}a/4\sim 3.5$~\AA. 

\begin{figure}
\begin{center}
\vspace{-15pt}
\includegraphics[scale=0.6]{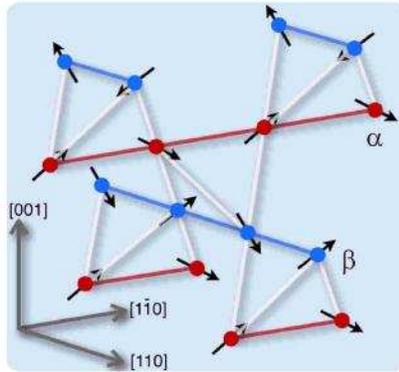}
\end{center}
\vspace{-15pt}
\caption{(Color online) $\alpha$ and $\beta$ chains in the pyrochlore lattice. 
From J.P.~Clancy {\em et al.}, Physical Review B {\bf 79}, 014408 (2009). http://link.aps.org/abstract/PRB/v79/p014408 
Copyright 2009 by the American Physical Society.}   
\label{alphabeta}
\end{figure}

It is also interesting to notice that the 
pyrochlore lattice for A sites can be considered as two sets of orthogonal 
chains, the one being parallel to [110] direction (called $\alpha$ chains) and 
the other one parallel to [1${\bar 1}$0] direction (refereed to $\beta$ chains, 
see Fig.~\ref{alphabeta}). 

\subsection{Single ion properties}

\begin{figure}
\begin{center}
\vspace{-15pt}
\includegraphics[scale=0.7]{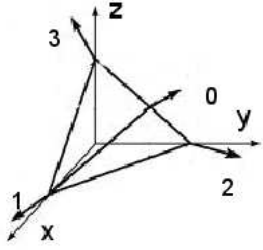}
\end{center}
\vspace{-15pt}
\caption{Elementary cell for the A-site pyrochlore oxide lattice. Vectors 
labeled 0,1,2,3 are unit vectors directed along [111] axes, which are 
distinguished by the magnetic ``easy-axis'' anisotropy.}   
\label{tetr}
\end{figure}

Dy (Ho) ions have [Xe]6S$^2$4f$^9$ (4f$^{10}$) ground state electron 
configuration. Rare-earth ions due to strong spin-orbit coupling form the 
total moment ${\bf J} = {\bf L} +{\bf S}$, where ${\bf L}$ (${\bf S}$) is the 
total orbital (spin) moment. According to Hund's rules, we can find $J=15/2$ 
for Dy$^{3+}$ ion with $L=3$ and $S=9/2$, and $J=8$ with $L=3$ and $S=5$ for 
Ho$^{3+}$. The $(2J+1)$ degeneracy of the configuration is lifted due to the 
crystalline electric field of ligands (oxygens). The crystal field Hamiltonian 
can be written for ${\bar 3}$m ($D_{3d}$) point symmetry of the A site 
as \cite{single,GinHerFau,YanGho,YanSenGho,MirBonHen}
\begin{eqnarray}
&&{\cal H}_{\rm cf} = \sqrt{4\pi}\sum_{l,m} {B_l^m Y_l^m\over \sqrt{2l+1}} = 
\sum_{l,m} B_l^m {\cal O}_l^m 
\nonumber \\
&&= B_2^0{\cal O}_2^0 +B_4^0{\cal O}_4^0 +B_4^3{\cal O}_4^3 
\nonumber \\
&&+B_6^0{\cal O}_6^0 +B_6^3{\cal O}_6^3 + B_6^6{\cal O}_6^6 \ , 
\end{eqnarray}
where $B_l^m$ are crystal field coefficients, $Y_l^m$ are spherical harmonics, 
and ${\cal O}_l^m$ are Stevens operators \cite{Stev,Hutch}, related to the 
projections of the total moment. For $L=3$ we limit ourselves with $l \le 6$, 
due to the Wigner-Eckart theorem. Further restriction comes about because the 
crystalline electric field environment is symmetric under operations of the 
point group $D_{3d}$. Here we use \cite{Stev,Hutch,Ryab,RudChu} 
${\cal O}_2^0 = 3J_z^2 -j(J+1)$, ${\cal O}_4^0 = 35J_z^4 -[30J(J+1)-25]J_z^2 
+3J^2(J+1)^2 -6J(J+1)$, ${\cal O}_4^{3}= (1/4)
[3(J_+^3-J_-^3)+2(J_+^3+J_-^3)J_z]$, where $J_{\pm}=J_x\pm iJ_y$, 
${\cal O}_6^0 = 231J_z^6 -[315J(J+1) -735]J_z^4 +[105J^2(J+1)^2 -525J(J+1) 
+294)J_z^2 -5J^3(J+1)^3 +40J^2(J+1)^2 -60J(J+1)$, ${\cal O}_6^6 = 
(1/2)(J_+^6+J_-^6)$, and ${\cal O}_6^{3}= (1/4)[3(40-3J(J+1))(J_+^3-J_-^3)+
(179-6J(J+1))(J_+^3+J_-^3)J_z + 99(J_+^3-J_-^3)J_z^2 +22(J_+^3-J_-^3)J_z^3]$.   
From the experiments on inelastic neutron scattering 
\cite{single,GinHerFau,YanGho,YanSenGho,MirBonHen} we can find 
that for Ho$_2$Ti$_2$o$_7$ (Dy$_2$Ti$_2$o$_7$) the ground state can be 
described as the Kramers doublet with $|J=8, m_{J}=\pm 8\rangle$ 
($|J=15/2, m_{J}=\pm 15/2\rangle$) with negligible contributions from 
components with other values of $m_J$ [higher multiplets are divided by the gap 
of order of $\Delta \sim 300$~K from the ground state Kramers doublet (in 
fact, the gap was estimated from 140~K to 380~K 
\cite{single,GinHerFau,YanGho,YanSenGho,MirBonHen})]. This is why, 
at low temperatures we can consider Ho$_2$Ti$_2$o$_7$ and Dy$_2$Ti$_2$o$_7$ as 
systems of effective Ising spins 1/2. However, the situation is different from 
the case of a standard uniaxial anisotropy, because most often the 
``easy axis'' is homogeneous for magnetic systems \cite{Ball,vV,AB1}, and in 
the considered case we have four equivalent ``easy axes'' for each tetrahedron 
parallel to [111] directions, see Fig.~\ref{tetr}.

We can introduce unit vectors directed along the ``easy axes'' 
${\bf e}_{0,1} = ({\bf x} \pm {\bf y} \pm {\bf z})/\sqrt{3}$, ${\bf e}_{2,3} = 
(-{\bf x} \pm {\bf y} \mp {\bf z})/\sqrt{3}$, where ${\bf x}$, ${\bf y}$ and 
${\bf z}$ are the unit vectors along the co-ordinate axes. Then at low 
temperatures $T \ll \Delta$ we can approximate 
\begin{equation}
{\bf J}_n \to |\langle J_z\rangle |\sigma_n^z {\bf e}_n \ , 
\end{equation}
where $\sigma^z_n$ are Pauli matrices, which have eigenvalues $\pm 1$, 
$|\langle J_z\rangle | \approx 15/2$ for Dy$_2$Ti$_2$O$_7$ and $|\langle 
J_z\rangle | \approx 8$ for Ho$_2$Ti$_2$O$_7$. There are no other components 
$\sigma_n^{x,y}$ in the low-temperature approximation, and, therefore, the 
low-energy physics can be approximated by the Ising model. Sometimes this 
situation is called classical, to stress that there is no spreading of 
excitations in the Ising model, and variables commute with each other. However, 
it can be misleading, because the Ising system has a discrete spectrum, the 
hallmark of quantum physics.

The external magnetic field ${\bf B}$ at low temperatures acts as 
\begin{eqnarray}
&&{\cal H}_Z = -g\mu_B \sum_n ({\bf B} \cdot {\bf J}_n) \nonumber \\
&&\approx -g\mu_B|\langle J_z\rangle | \sum_n ({\bf B} \cdot {\bf e}_n)
\sigma^z_n \ , 
\end{eqnarray}
where $\mu_B$ is Bohr's magneton, equal to 9.27$\times 10^{-24}$~J/T, or, more 
convenient for us, 0.671~K/T, if we measure all energies in Kelvins, and $g$ 
is the Land\'e $g$-factor, equal to 4/3 for Dy$^{3+}$ and 5/4 for Ho$^{3+}$. 
Then the characteristic energy scale for the Zeeman interaction of pyrochlore 
oxides is $10\mu_B |B| \sim 6.71 |B|$~K, where the magnitude of the 
magnetic field $|B|$ is measured in Tesla. Obviously, for the values 
of the field $|B| \le 45$~T such an energy is lower than $\Delta \sim 300$~K, 
and the Ising approximation is justified. For higher values of the field 
we have to take into account higher-energy multiplets due to crystalline 
electric field. 

We can also neglect the van Vleck contribution to the magnetic susceptibility 
of the considered pyrochlore oxides, and the contributions from the high-energy 
multiplets, caused by the crystalline electric field, to the susceptibility at 
low temperatures. 
  
\subsection{Realization of the ice rule} 

\begin{figure}
\begin{center}
\vspace{-15pt}
\includegraphics[scale=0.6]{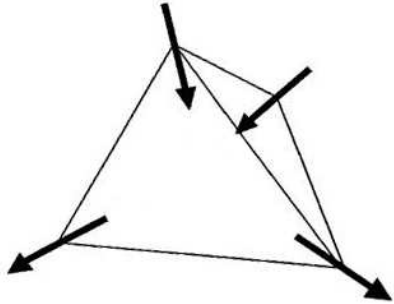}
\end{center}
\vspace{-15pt}
\caption{Realization of the ice rule in the primitive cell of a spin ice.}   
\label{icerule}
\end{figure}

Now we are in position to explain why these pyrochlore oxides 
are known as spin ices. Namely, let us consider the Hamiltonian of exchange 
interactions between rare-earth magnetic ions ${\cal H}_{ex} = -(1/2)\sum_{i,j} 
{\cal J}_{i,j} ({\bf J}_i\cdot {\bf J}_j)$, where $i$ and $j$ denote positions 
of magnetic ions, and ${\cal J}_{i,j}$ are the exchange integrals. The 
prefactor (1/2) is introduced to avoid double counting of sites. Here we limit 
ourselves with the isotropic version of the exchange coupling. Notice, 
however, that the symmetry allows four distinct types of anisotropic exchange 
interactions in a pyrochlore lattice \cite{exch,McCCurGin}. For rare-earth 
systems 4f orbitals are screened by 5s and 6p orbitals, and, therefore, the 
exchange interaction (both, the direct exchange between rare-earth ions 
themselves, and the indirect exchange via oxygen ions O$^{2-}$) is expected to 
be small. Then at low temperatures we can approximate that expression as
\begin{equation}
{\cal H}_{ex} \approx - |\langle J_z\rangle|^2 \sum_{i,j}{\cal J}_{i,j}
({\bf e}_i\cdot {\bf e}_j) \sigma_i^z\sigma_j^z \ . 
\label{He}
\end{equation}
The value of the exchange integrals for nearest neighboring Dy$^{3+}$ in 
Dy$_2$Ti$_2$O$_7$ has been estimated \cite{HG,exch1} as ${\cal J}_{ij} \sim 
0.66$~mK. We can introduce $J \equiv |\langle J_z \rangle|^2{\cal J}_{i,j}$. 
For Ho-based titanate its value is estimated as $J \sim 4.22$~K, and for 
Dy-based titanate it is $J \sim 3.71$~K, i.e., in both cases $J \ll \Delta$. 
Notice that $J >0$, i.e. it corresponds to the {\em ferromagnetic} nearest 
neighbor coupling in the initial exchange Hamiltonian. For nearest neighbors 
we can write cf. \cite{M98}
\begin{equation}
{\cal H}_{ex} \approx -J \sum_{\langle i,j\rangle} 
({\bf e}_i\cdot {\bf e}_j) \sigma^z_i\sigma^z_j = (J/3)\sum_{\langle i,j\rangle} 
\sigma^z_i\sigma^z_j \ , 
\end{equation}
where we limit ourselves with the nearest neighbors $\langle i,j\rangle$, and 
used the equality $({\bf e}_i\cdot {\bf e}_j)=-1/3$ for the tetrahedron, 
which follows from the definition of the unite vectors ${\bf e}_n$. Using the 
definition $S_t = (\sigma^z_0 +\sigma^z_1 +\sigma^z_2 +\sigma_z^3)_t$ as the 
total effective Ising spin for the tetrahedron, we can get \cite{MC}
\begin{equation}
{\cal H}_{ex} \approx {J\over 6} \sum_t S_t^2 - {2N_t J\over 3} \ , 
\label{tetrex}
\end{equation}
where the summation is over each tetrahedron primitive cells, $N_t$ is the 
total number of such cells. It is important to emphasize that the 
ferromagnetic exchange between real total moments in such pyrochlore oxides 
produces the effective {\em antiferromagnetic} coupling between effective Ising 
spins. Then, it is clear from Eq.~(\ref{tetrex}) that for $J>0$ the lowest 
energy has the state with $S_t=0$. It is equivalent to the ice rule in the 
water ice, or to the Verwey's tetrahedron rule: The lowest energy is related 
to states of each tetrahedron with to Ising spins directed inside, and two 
others directed outside the tetrahedron which means two of $\sigma^z_n$ have 
+1 eigenvalue, and two others have -1 eigenvalue (notice that all spins have 
directions parallel to [111] axes, cf. Fig.~\ref{icerule}). The fulfillment of 
the ice rule implies the frustration, totally equivalent to the water ice 
I$_{\rm h}$. That suggested the name for such systems: Spin ices! It is 
interesting to remark that the antiferromagnetic exchange between total 
moments has to manifest the non-frustrated state with all effective Ising 
spins having the same signs of their eigenvalues. In other words, the real 
antiferromagnetic exchange produces the ground state with all effective Ising 
spins directed either in or outside each tetrahedron.    

At low temperatures additional spin-spin interaction may manifest itself, 
the magnetic dipole-dipole interaction, which Hamiltonian is (the importance 
of the magnetic dipole-dipole interactions for spin ices has been pointed out, 
e.g., in \cite{Sha})
\begin{eqnarray}
&&{\cal H}_{d} = \left[{\mu_0\over 4\pi}\right] {g^2\mu^2_{\rm B}\over 2x^3} 
\sum_{i,j} \biggl[ {({\bf J}_i\cdot {\bf J}_j)\over (|{\bf r}_{ij}|^3/x^3)} 
\nonumber \\
&&- {3({\bf r}_{ij}\cdot {\bf J}_i)
({\bf r}_{ij}\cdot {\bf J}_j)\over (|{\bf r}_{ij}|^5/x^3)} \biggr]
\nonumber \\
&&\approx \left[{\mu_0\over 4\pi}\right] {g^2\mu^2_{\rm B}|\langle J_z 
\rangle|^2 \over 2x^3} \sum_{i,j} \sigma_i^z\sigma_j^z
{({\bf e}_i\cdot {\bf e}_j)\over (|{\bf r}_{ij}|^3/x^3)} \nonumber \\
&&- {3({\bf r}_{ij}\cdot {\bf e}_i)({\bf r}_{ij}\cdot {\bf e}_j) \over 
(|{\bf r}_{ij}|^5/x^3)} 
\label{Hd}
\end{eqnarray} 
where ${\bf r}_{ij} = {\bf r}_i -{\bf r}_j$, and $\mu_0/4\pi = 10^{-7}$~N~A
$^{-2}$ ($\mu_0$ is the vacuum permeability), and for convenience we have 
normalized every contribution to the value for the nearest neighbors. Then 
the magnetic dipole-dipole contribution to the nearest neighbor interaction 
between effective Ising spins in the primitive cell is $D = 
\mu_0g^2\mu^2_{\rm B}|\langle J_z \rangle|^2/ 8\pi x^3$. We can estimate this 
value as 1.4~K. Hence, the nearest neighbor part of the magnetic 
dipole-dipole interaction renormalizes the exchange interaction \cite{HG,B1}, 
and the conclusions made above using the only exchange coupling seem to hold 
for the nearest neighbor dipole-dipole interactions. However, it is not the 
total story. Unlike exchange couplings, magnetic dipole-dipole interactions 
are long-ranged (the model, which take into account long-ranged dipole-dipole 
coupling is called the dipole spin ice). This has been taken into account in 
several ways. First, some truncation for the dipole-dipole interaction was 
used \cite{HG,Sid}. Then Ewald summation \cite{Ew}, usually used for the 
estimation of the long-range dipole-dipole coupling \cite{MG}, Monte-Carlo 
simulations \cite{MG,BarNew,MHG,RufMelGin}, and mean-field calculations 
\cite{mf,GinHer1,IsMoSon} were also performed. They found that the long-range 
part of the magnetic dipole-dipole interaction can produce the long-range 
N\'eel-like ordering (via the first order phase transition) at very low 
temperatures. Hence, the dipolar spin ice is characterized by ordering with 
the commensurate propagation wave vector of the order parameter, and, 
therefore, the long-range dipole-dipole coupling removes the degeneracy, 
caused by the frustration. From this perspective, spin ices are equivalent to 
the water ice, which manifests the transition from the frustrated hexagonal 
and cubic phases to the orthorhombic ice XI phase. Such a phase transition in 
spin ices has not been observed yet see, e.g., \cite{Fuz}. From the 
theoretical viewpoint, it was unclear why the ice rule is satisfied in dipolar 
spin ices. The reason has been clarified in \cite{Is1}. The autors of 
\cite{Is1} have pointed out that the tetrahedron (ice) rule $S_t=0$, or more 
general, $\sum_i S_{t,i} =0$, where the sum is performed over all tetrahedra, 
is equivalent to some ``spin field'' ${\bf B}_s({\bf r})$, which divergence is 
zero, $(\nabla \cdot {\bf B}_s)=0$ (notice that we deal with the lattice 
case). Nonzero flux of that field means broken tetrahedron (ice) rule. Then 
the correlation functions for the field are 

\begin{equation}
\langle B_{\alpha}({\bf r})B_{\beta}(0)\rangle \sim {3x_{\alpha}x_{\beta} -
r^2\delta_{\alpha,\beta}\over r^5} \ , 
\end{equation}
where $\alpha,\beta =x,y,z$. The dipolar Hamiltonian on the pyrochlore lattice 
can be presented as the projector, and the correlations of the projector are 
equivalent to the projector itself \cite{Is1}. That means that the local 
constraint, i.e., the tetrahedron rules, yield dipolar-like correlations at 
large distances. these Coulomb correlations are the signature of the 
so-called ~Coulomb phase'' in dipole spin ices, see, e.g. 
\cite{Coul,IsMoSon,KheMoParSon,FDWSPBAMB}. 

\subsection{Experimental discovery of spin ices}

In fact, spin ices were discovered experimentally in \cite{H1,Har,BraGing,R1}. 
The authors 
of \cite{H1} (who, actually, first used the term spin ice) performed the 
neutron scattering experiment to investigate low (down to 0.05~K) temperatures 
properties of Ho$_2$Ti$_2$O$_7$. At zero magnetic field they have 
observed no magnetic ordering by the neutron scattering (down to 0.35~K) and 
by muon spin rotation (down to 0.05~K). They have found positive 
$\theta_{\rm CW} \approx 1.9$~K, indicating ferromagnetic interactions. It is 
interesting that $\theta_{\rm CW}$ is of the same order of values as both $D$ 
and $J$. naturally, the absence of $T_c$ (or, similar, $T_c \ll 
\theta_{\rm CW}$) implies very high level of frustration in this compound. The 
magnetic field affected the neutron scattering depending on the pre-history of 
the sample (see also \cite{Kan,Kad,Pet1}). It is similar to the situation in 
spin glasses \cite{sg,EdAnd,SherKirk,MezParVir,FishHer,Myd,BY}, despite absent 
(or very weak) randomness in Ho$_2$Ti$_2$O$_7$. Such experiments have been 
tried to be explained \cite{BH} by considering the ferromagnetic Ising spins 
directed along [111] in the pyrochlore lattice. 

\begin{figure}
\begin{center}
\vspace{-15pt}
\includegraphics[scale=0.5]{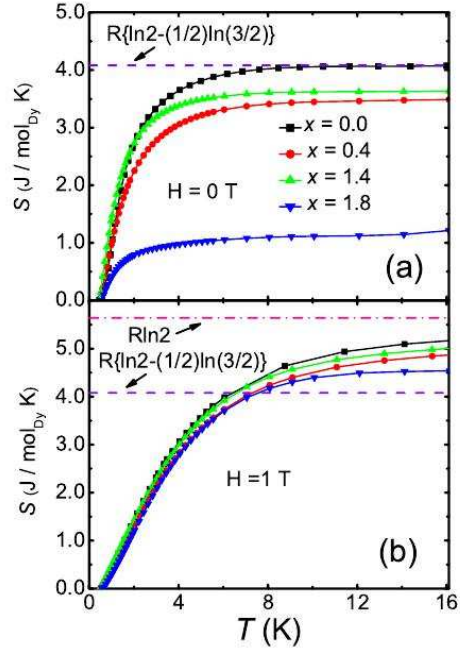}
\end{center}
\vspace{-15pt}
\caption{(Color online) Entropy of Dy$_{2-x}$Y$_x$Ti$_2$O$_7$ as a function of 
temperature without (upper panel)  and with the external magnetic field. We 
can see that the entropy is related to magnetic ions Dy$^{3+}$, because higher 
concentration of Y$^{3+}$ ions decreases the value of ${\cal S}$ in the system. 
From X.~Ke {\em et al.}, Physical Review Letters {\bf 99}, 137203 (2007). http://link.aps.org/abstract/PRL/v99/p137203  
Copyright 2007 by the American Physical Society.}   
\label{entropy}
\end{figure}

Even more important for the analogy between the hexagonal water ice and the 
spin was the measurement of the specific heat $c(T)$ in Ref.~\cite{R1} in 
Dy$_2$Ti$_2$O$_7$. The authors of \cite{R1} followed the strategy of 
Ref.~\cite{Gia,Gia1} to get the value of the residual entropy in spin ice 
material, similar to what had been performed for the water ice. The entropy 
for the water ice has been calculated by integration of $c(T)/T$ starting from 
10~K till the gas phase, 
\begin{equation}
\Delta {\cal S}_{1,2} = {\cal S}(T_1)-{\cal S}(T_2) = \int_{T_1}^{T_2} 
{c(T) dT \over T} \ , 
\end{equation}
with an addition of the latent  heat at the melting and vaporization. Then, 
calculated in such a way value has been compared with the expected from 
calculations (see above) absolute value for the hexagonal water ice. 
In Ref.~\cite{R1} it has been measured the magnetic specific heat of 
Dy$_2$Ti$_2$O$_7$ between the temperatures $T_1=0.3$~K and $T_2 =10$~K. The 
former, according to estimations, is related to the spin ice phase (its value 
was lower than the 
temperature of the maximum in the dependence $c(T)$ at 1.24~K, identified as 
the crossover temperature \cite{R1}), while the latter expected to 
be already in the paramagnetic regime, where one can neglect spin-spin 
interactions. The Schottky-like maximum in $c(T)$ was connected to the energy 
gap between the states satisfying the ice rule (two spins directed ``in'' the 
tetrahedron, and two ``out'') and the excited state with one spin ``in'' and 
three spins directed ``out'' (or vice versa). In the low temperature regime, 
$T < 1.24$~K, the spin flip rate has been calculated to be exponentially 
decaying \cite{MG}, because of the steady-state settling to the ice rule 
``phase''. The restored that way value of the magnetic entropy in the limit 
$T \to T_2$ was obtained as 3.9~J~mol$^{-1}$K$^{-1}$, which is very close to 
the difference between the magnetic entropy in the paramagnetic regime (free 
effective Ising spins), $k_{\rm B}N_{\rm A}\ln (2) = 5.76$~J~mol$^{-1}$K$^{-1}$ 
(here $N_{\rm A}$ is the Avogadro number), and the Pauling value 
$k_{\rm B}N_{\rm A}\ln \sqrt{3/2} = 1.68$~J~mol$^{-1}$K$^{-1}$. The results of 
later experiments \cite{Higa,Hir,Ke} have shown even better agreement between 
the measured value of the entropy and the one, predicted by Pauling, see 
Fig.~\ref{entropy}. Moreover, by studying Dy$_{2-x}$Y$_x$Ti$_2$O$_7$, i.e., 
by replacing the magnetic Dy$^{3+}$ ion by the non-magnetic Y$^{3+}$, 
\cite{Ke} has proven that the considered entropy is related namely to magnetic 
subsystem of spin ices. Performed Monte-Carlo simulations \cite{HG} show a very 
good agreement with the observed temperature behavior of the specific heat and 
entropy in Dy$_2$Ti$_2$O$_7$. 

On the other hand, Ho$_2$Ti$_2$O$_7$ does not show such a direct evidence of 
the low temperature specific heat and entropy, as Dy$_2$Ti$_2$O$_7$. The 
reason for such a difference in the behaviors of two representatives of the 
spin ice group is connected with the anomalously large hyperfine interaction 
between nuclear and electron spins, characteristic for Ho$^{3+}$. Such an 
interaction manifests itself in a Schottky anomaly in the temperature 
dependence of the magnetic specific heat of Ho$_2$Ti$_2$O$_7$ at about 0.3~K. 
If one subtracts the nuclear contribution, the residual Pauling entropy in 
Ho$_2$Ti$_2$O$7$ can be manifested \cite{CG,B1}. Similar observations 
\cite{Kad,Ke1,Mats,Mats1} were made for stannates Ho$_2$Sn$_2$O$_7$ and 
Dy$_2$Sn$_2$O$_7$.

Magnetic Coulomb phase in spin ices has been observed in \cite{FDWSPBAMB} via 
the polarized neutron scattering.

\subsection{Spin ices in an external magnetic field} 

First magnetic measurements in Dy$_2$Ti$_2$O$_7$ have been performed in  
\cite{Blo,Flood}, in which the dc magnetic susceptibility and magnetic moment 
have shown a strong magnetic (Ising) anisotropy along [111].

The temperature and magnetic field dependencies of thermodynamic 
characteristics of spin ice systems can be calculated in the simplest way, 
e.g., by performing recently developed approach \cite{Tim}, in which the 
Bethe-Peierls approximation on a Bethe lattice was used. It assumes that 
effective fields acting on effective Ising spins of the primitive cell, 
Fig.~\ref{tetr}, are the same as the ones, acting on cell's nearest neighbors, 
which does not, actually, hold in pyrochlore system. However, results, 
obtained in this approximation, show a good qualitative agreement with the 
experimentally observed data, see below. In the Bethe-Peierls approach the 
free energy of the spin-ice system per rare-earth ion can be written as 
\begin{eqnarray} 
&&F= {k_{\rm B}T\over 4} \sum_{n=0}^{3} \ln [2\cosh(2f_{n} -b_{n})] 
\nonumber \\
&&-{k_{\rm B}T\over 2} \ln [2Z({\bf f})] \ , 
\label{F} 
\end{eqnarray}
where  
\begin{equation}
Z({\bf f}) = \sum_{m=0}^2 Z_m({\bf f})e^{-2m^2J/k_{\rm B}T}. 
\label{Z}
\end{equation}
The index $n$ denotes four directions (cf. Fig.~\ref{tetr}) for the 
``easy axes'' of the magnetic anisotropy (considered here to be much larger 
than the effective interactions between effective Ising spins) in each 
tetrahedron in the spin-ice system, see above, $b_{n} \equiv 
g\mu_{\rm B}|\langle J_z\rangle|({\bf e}_{n} \cdot {\bf B})/k_{\rm B}T$ are the 
projections of the external magnetic field ${\bf B}$ normalized by the 
temperature, and $f_{n}$ are the projections of the effective magnetic field, 
which acts on the effective Ising spins in the considered tetrahedron from 
other effective Ising spins in the system. In Eq.~(\ref{Z}) $J$ denotes the 
value of the effective exchange interaction between spins in each tetrahedron, 
and 
\begin{eqnarray} 
&&Z_0({\bf f}) = \cosh (f_0 +f_1-f_3-f_3) 
\nonumber \\
&&+2\cosh(f_0-f_1)\cosh(f_2-f_3) \ , 
\nonumber \\
&&Z_1({\bf f}) = \sum_{n=0}^3 \cosh\left( \sum_{k=0}^3 f_{k} 
-2f_{n} \right) \ , \nonumber \\
&&Z_2 = \cosh \left( \sum_{n=0}^3 f_{n} \right) \ . 
\label{Zn}
\end{eqnarray}
are related to the three possible spin configuration in each tetrahedron: 
two spins directed inside tetrahedron and two spins directed outside (
``two in and two out''); ``three in and one out'' (or vice versa, ``one in and 
three out''); and ``four in'' (or ``four out''), so that for larger $J$ the 
most favorable configuration in the absence of the external field is ``two in 
and two out''. It turns out that the sign of $J$ in the approach is taken 
such that ``two in and two out'' configuration has the lowest energy. The 
value of the  effective exchange constant $J$ can be chosen to satisfy the
experimental data in spin ice systems. Basically, we consider the spin ice 
model (with the nearest neighbor exchange coupling between effective Ising 
spins) in the external magnetic field, i.e., we study the low-temperature 
Hamiltonian ${\cal H}_{ex} +{\cal H}_Z$ in the Bethe-Peierls approximation. 
Notice, however, that the mean-field-like Bethe-Peierls approximation becomes 
better for long-range interactions, i.e., it can be applied to dipole spin 
ices as well.  The values of the projections of the effective field $f_{n}$ 
satisfy the following set of equations 
\begin{equation} 
\tanh (2f_{n} -b_{n}) = {\partial \ln Z({\bf f}) \over 
\partial f_{n}}  \ .    
\label{mean}
\end{equation}
The value of the average effective Ising spin moment (related to the 
low-temperature magnetization per rare earth ion divided by 
$g\mu_{\rm B}|\langle J_z\rangle |$) in this approximation can be written as 
\begin{equation}
M = {1\over 4} \sum_{n=0}^3 {\bf e}_{n} 
\tanh (2f_{n} -b_{n}) \ , 
\label{S0}
\end{equation}
and the magnetic susceptibility is 
\begin{equation}
\chi_0 = {\partial M \over \partial B} \ .
\label{susc}
\end{equation}
The entropy per rare-earth ion can be written as 
${\cal S} = -\partial F/ \partial T$, and the specific heat is 
$c =T(\partial S/ \partial T)$.  

Let us consider three directions of the magnetic field, namely ${\bf B} 
\parallel [111]$, ${\bf B} \parallel [100]$ and 
${\bf B} \parallel [011]$. In the first direction ${\bf B} = 
B{\bf e}_0$, so that $b_0= g\mu_{\rm B}|\langle J_z\rangle| 
B/k_{\rm B}T$, $b_{1,2,3} = -g\mu_{\rm b}|\langle J_z\rangle| B /3k_{\rm B}T$. For 
the second direction of the field we have ${\bf B} = B{\bf x}$, and $b_{0,1} 
=g\mu_{\rm B}|\langle J_z \rangle |B/\sqrt{3}k_{\rm B}T$, $b_{2,3}=-g\mu_{\rm B} 
|\langle J_z\rangle |B/\sqrt{3}k_{\rm B}T$. For the third direction ${\bf B} = 
B({\bf x} +{\bf y})/\sqrt{2}$, i.e., $b_0=-b_1=\sqrt{2/3}g\mu_{\rm B}
|\langle J_z\rangle| B/k_{\rm B}T$, and $b_2=b_3=0$. 

It is important to notice that for all field directions the results depend 
\cite{Tim} on the order of limitations $T \to 0$ and $B \to 0$ (``field 
cooling'' or ``zero field cooling''). Let us start with $B \to 0$ case, i.e., 
``zero field cooling''. For such a condition we have $M =0$ and 
$\chi = 2(g\mu_{\rm B}|\langle J_z\rangle |)^2/3k_{\rm B}T$, with the Pauling 
value of the remnant entropy, as it must be, and $c=0$.   

\begin{figure}
\begin{center}
\vspace{-15pt}
\includegraphics[scale=0.35]{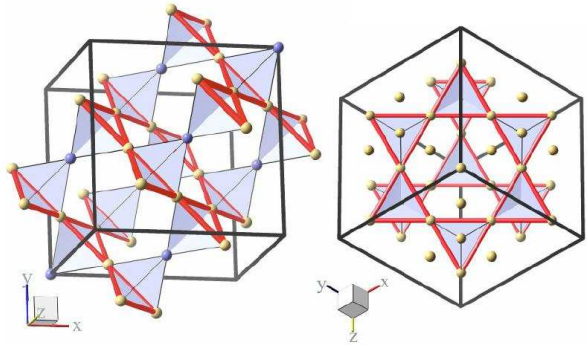}
\end{center}
\vspace{-15pt}
\caption{(Color online) Left panel: A-sites of the pyrochlore lattice. Right 
panel: Kagome lattice formed by layers perpendicular to [111]. From 
K.A.~Ross {\em et al.}, Physical Review Letters {\bf 103}, 227202 (2009). http://link.aps.org/abstract/PRL/v103/p227202  
Copyright 2009 by the American Physical Society.}   
\label{kagom-pyr}
\end{figure}

The ``field cooled'' case, $T \to 0$ first, implies for the field directed 
along [111],  $\chi=c=0$ and $M=1/3$, with ${\cal S} = 
(1/4)k_{\rm B}\ln (4/3)$. The entropy is reduced with respect to the Pauling 
value, because the ground state degeneracy is partly lifted due to the field 
directed along [111]. Such a field fixes the direction of the effective spin 
with the index 0, while three others are free. Such a phase is related to 
the ``Kagome ice'' state of the pyrochlore lattice, see Fig.~\ref{kagom-pyr}.   
Such a reduction of the entropy due to the external field directed along [111] 
has been observed \cite{HMTTS,HFM} in Dy$_2$Ti$_2$O$_7$.  
   
\begin{figure}
\begin{center}
\vspace{-15pt}
\includegraphics[scale=0.2]{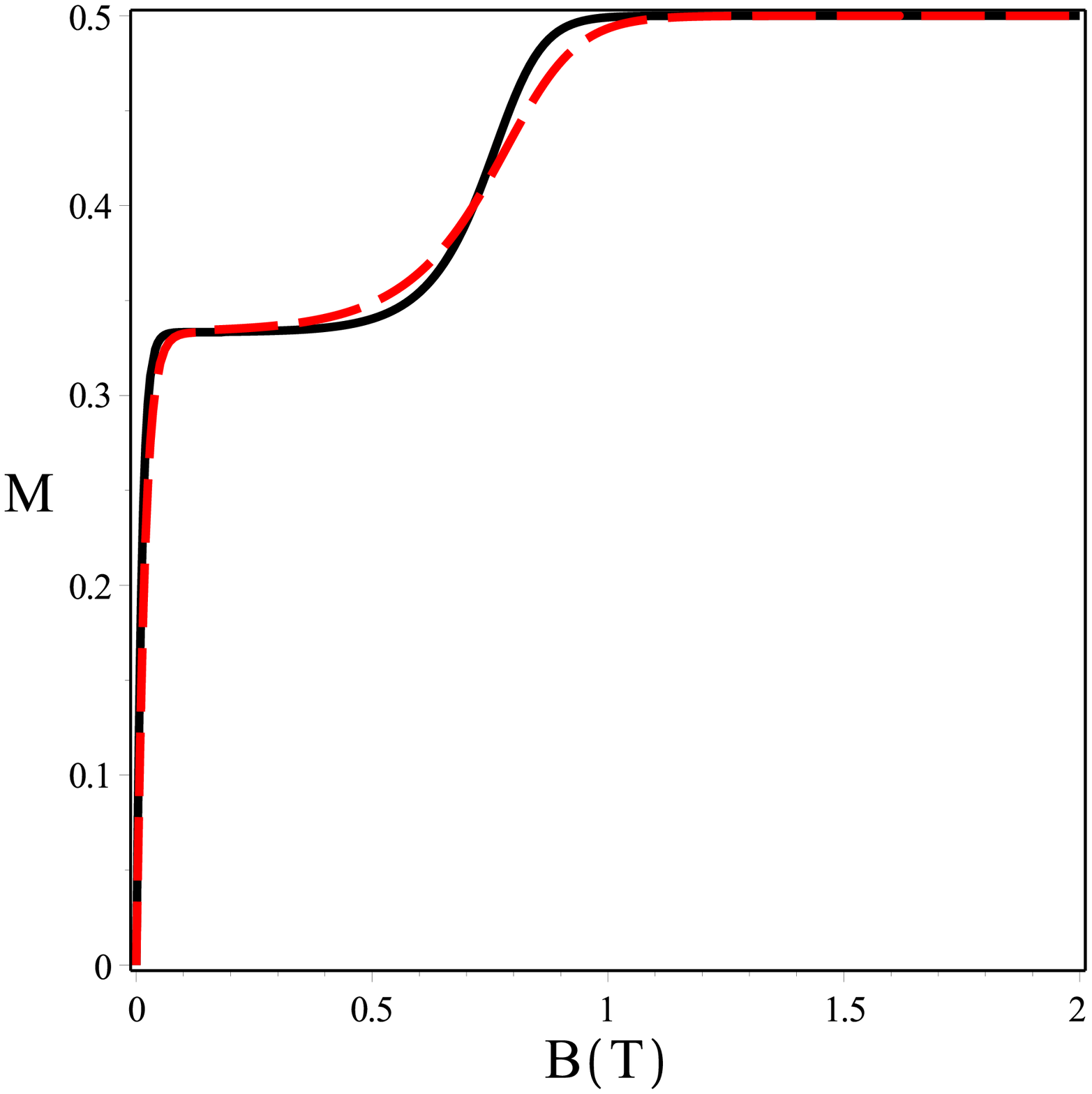}
\includegraphics[scale=0.2]{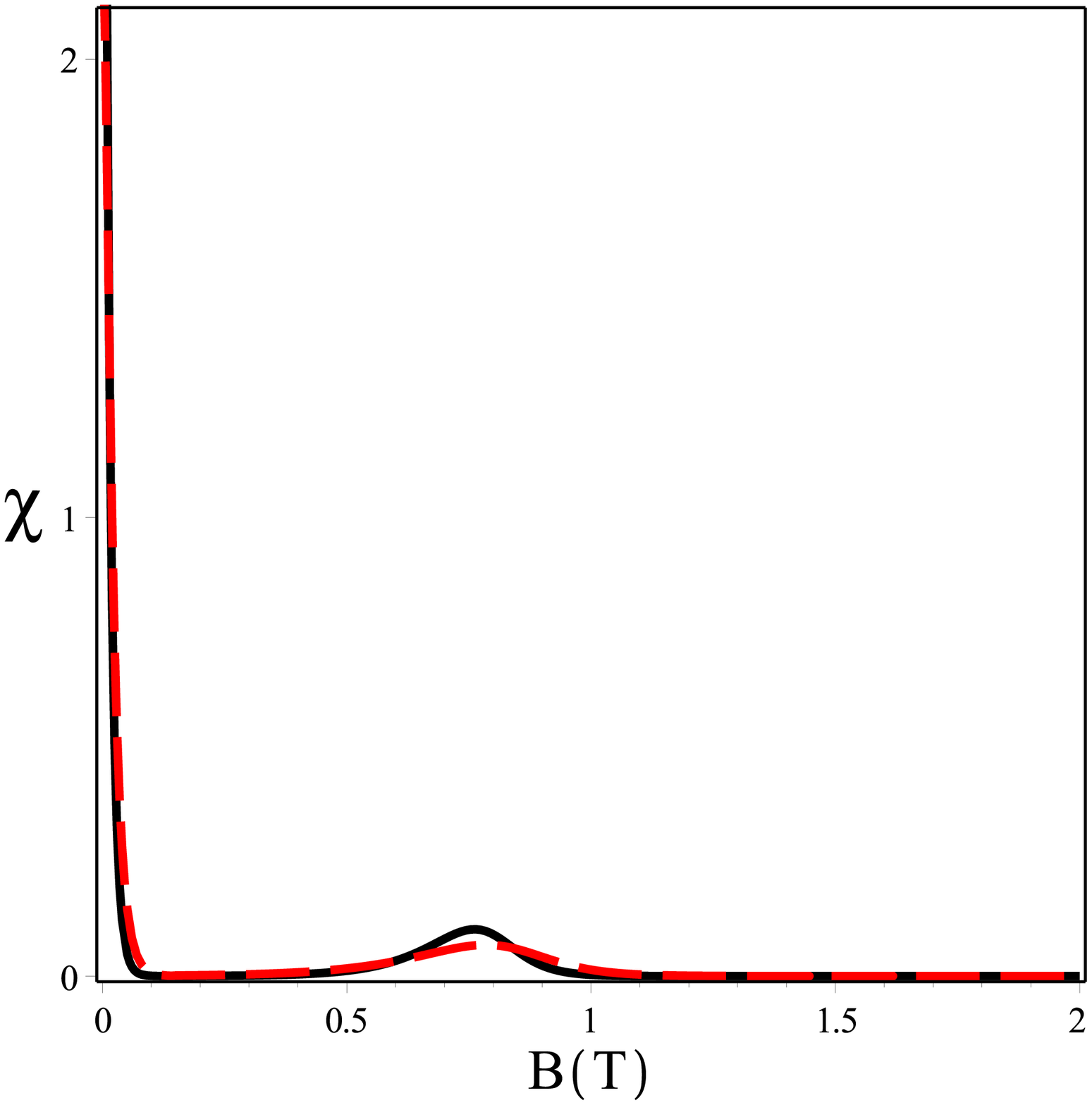}
\end{center}
\vspace{-15pt}
\caption{(Color online) The average effective spin moment per rare earth ion 
$M$ (i.e., it is the magnetization per rare earth ion divided by 
$g\mu_{\rm B}|\langle J_z \rangle|$), right panel, and the magnetic 
susceptibility $\chi$ (divided by $g\mu_{\rm B}|\langle J_z 
\rangle|/k_{\rm B}$), left panel, as a function of the external field $B$ 
parallel [111] [for $J=0.8$~K and $T=0.2$~K (solid black curves), and for 
$T=0.3$~K (dashed red curves) calculated within the Bethe-Peierls 
approximation for the spin ice model.]}    
\label{m111}
\end{figure}

At the critical values of the external magnetic field, directed along [111], 
the field behavior of the average spin moment shows the jump-like features 
from the state with the value of the spin moment zero to the value with the 
average moment 1/3 of the nominal one (1), and then, to the state with the 
value 1/2 of the nominal (that jump takes place at $g\mu_{\rm B}|\langle 
J_z\rangle |B_c = 6J$), see Fig.~\ref{m111}. 

The growing temperature ``smears out'' those features, slightly shifting the 
positions of them to higher values of the field. The features near $B =0$ are 
related to the step-like feature of the magnetic field behavior of the 
magnetization. It is, in fact, the the manifestation of the transition between 
the spin ice and Kagome ice phases in the external magnetic field directed 
along [111].  For other approach to the calculation of magnetization in spin 
ices in the external [111]-directed magnetic field consult \cite{IRMS}. 

\begin{figure}
\begin{center}
\vspace{-15pt}
\includegraphics[scale=0.3]{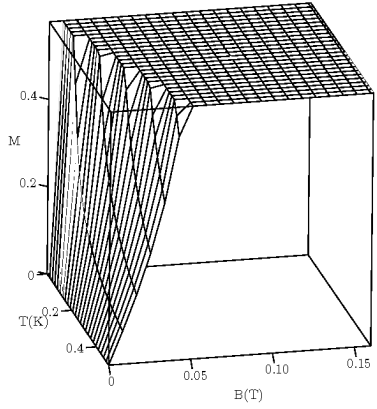}
\end{center}
\vspace{-15pt}
\caption{The average effective spin moment per rare earth ion $M$ as a 
function of temperature and external magnetic field $B$ directed along [100], 
calculated for the spin ice model within the Bethe-Peierls approximation. The 
small features on the surface are artifacts of the numerical 
computation/drawing.}    
\label{m100}
\end{figure}

\begin{figure}
\begin{center}
\vspace{-15pt}
\includegraphics[scale=0.35]{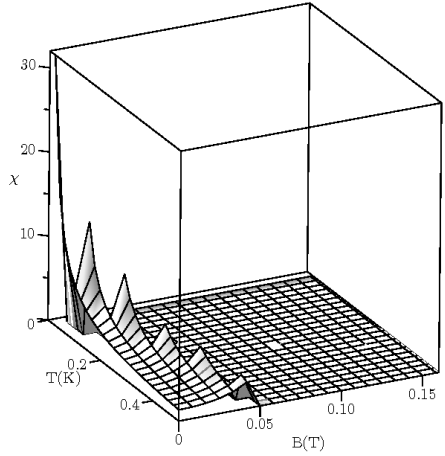}
\end{center}
\vspace{-15pt}
\caption{The magnetic susceptibility per rare earth ion $\chi$ (divided by 
$g^2\mu^2_{\rm B}|\langle J_z\rangle|^2/k_{\rm B}$) as a function of temperature 
and external magnetic field $B$ directed along [100], calculated for the spin 
ice model within the Bethe-Peierls approximation. The small features on the 
surface are artifacts of the numerical computation/drawing.}    
\label{chi100}
\end{figure}

\begin{figure}
\begin{center}
\vspace{-15pt}
\includegraphics[scale=0.35]{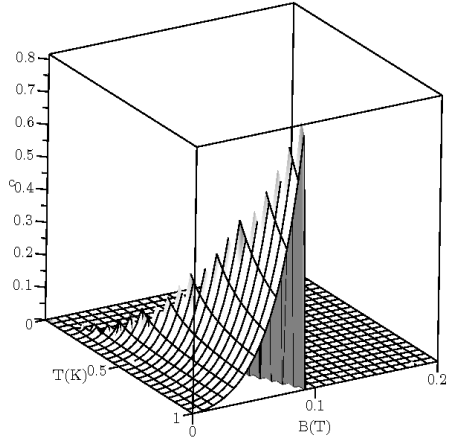}
\end{center}
\vspace{-15pt}
\caption{The magnetic contribution to the specific heat per rare earth ion $c$ 
as a function of temperature and external magnetic field $B$ directed along 
[100], calculated for the spin ice model within the Bethe-Peierls 
approximation. The small features on the surface are artifacts of the numerical 
computation/drawing.}    
\label{c100}
\end{figure}

Now let us consider the [100] direction of the applied magnetic field. Here 
the solution of Eqs.~(\ref{mean}) implies the following behavior for the 
characteristics of the spin-ice system. In the ``field-cooled'' case the 
degeneracy is completely lifted and at $T=0$ we should have ${\cal S} =0$. 
Then the increase of the value of the field results in the 
``pseudo-transition'' \cite{MoSon,Kast1} from the spin ice state to the 
``saturated'' state. Indeed at $B=B_K = \sqrt(3) \ln(2)k_{\rm B}T
/2g\mu_{\rm B}|\langle J_z \rangle |$, a Kasteleyn transition \cite{Kast,Kasta}, 
first predicted for the model of dimers on a two-dimensional lattice, takes 
place from the spin ice phase with the Pauling residual entropy to the 
``saturated'' state with zero entropy and the average effective spin moment 
a little larger than 1/2 of the nominal value. It happens because one of six 
spin configurations of ``two in - two out'' spin ice becomes preferable in 
such a field, which completely lifts the ground state degeneracy. Notice that 
this transition is seen in the external magnetic field at the temperature 
$T=T_K = 2g\mu_{\rm B}|\langle J_z\rangle |B /k_{\rm B}\sqrt(3)\ln(2)$: For 
$T < T_K$ the average effective spin moment is about 1/2 of the nominal value, 
see Fig.~\ref{m100}, and the magnetic susceptibility is zero, see 
Fig.~\ref{chi100}. At $T=T_k$ the temperature dependence of the magnetic 
susceptibility shows a jump-like feature (a cusp in the temperature behavior 
of the average effective spin moment), and for $T > T_K$ both the average 
moment and the magnetic susceptibility decay with the growth of temperature. 
The magnetic contribution to the specific heat also manifests features at the 
Kasteleyn-like transition in its temperature and magnetic field behavior, see 
Fig.~\ref{c100}. Such a transition has been observed, e.g., \cite{FBMMW} in 
Ho$_2$Ti$_2$O$_7$. 

\begin{figure}
\begin{center}
\vspace{-15pt}
\includegraphics[scale=0.3]{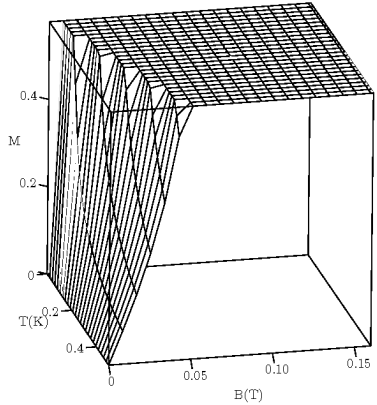}
\end{center}
\vspace{-15pt}
\caption{The magnetic moment per rare earth ion $M$ divided by 
$g^2\mu^2_{\rm B}|\langle J_z\rangle|^2/k_{\rm B}$ as a function of temperature 
and external magnetic field $B$ directed along [011], calculated for the spin 
ice model within the Bethe-Peierls approximation.}   
\label{m011}
\end{figure}

\begin{figure}
\begin{center}
\vspace{-15pt}
\includegraphics[scale=0.4]{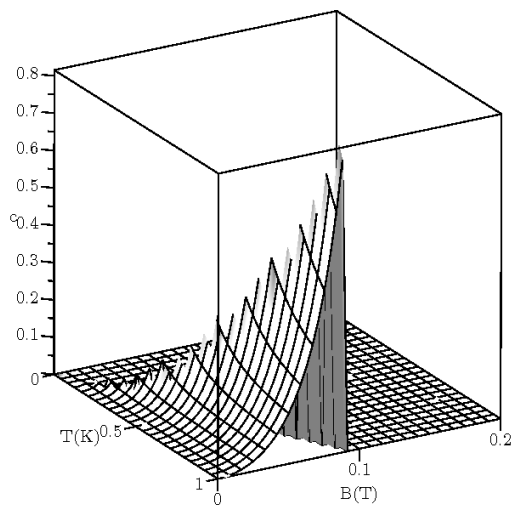}
\end{center}
\vspace{-15pt}
\caption{The magnetic susceptibility per rare earth ion $\chi$ (divided by 
$g^2\mu^2_{\rm B}|\langle J_z\rangle|^2/k_{\rm B}$) as a function of temperature 
and external magnetic field $B$ directed along [011], calculated for the spin 
ice model within the Bethe-Peierls approximation.}   
\label{chi011}
\end{figure}

The field along [011] does not affect effective Ising spins 2 and 3, which 
form $\beta$ chains, and act only on spins, belonging to $\alpha$ chains, see 
Fig.~\ref{alphabeta}. In the ground state for the ``field cooled'' case the 
spin at the position 0 is directed ``out'', and the spin at the position 1 is 
directed ``in'', while directions of effective spins from the $\beta$ chain 
are not fixed. We have at $T=0$ for the ``field cooled'' case ${\cal S} =c =
\chi=0$ and $M=g\mu_{\rm B}|\langle J_z\rangle|/\sqrt{6} \approx 0.41 
g\mu_{\rm B}|\langle J_z\rangle|$. The results of calculations of the 
temperature and magnetic field dependences of the average effective Ising spin 
and magnetic susceptibility for [011] direction of the field are shown in 
Figs.~\ref{m011} and \ref{chi011}. The crossover between the spin ice state 
and ``ordered chain'' states has been also observed experimentally 
\cite{HMO,YosNeWa}. 

The behavior of thermodynamic characteristics of the spin ice model for other 
directions of the external field can be calculated in a similar way \cite{Tim}.
It is important to point out that at nonzero values of the field the 
magnetization of spin ice systems becomes essentially nonzero, which implies 
the necessity to take into account demagnetization factors of the samples, 
when comparing theoretical results with the experimentally observed data. 

\begin{figure}
\begin{center}
\vspace{-15pt}
\includegraphics[scale=0.7]{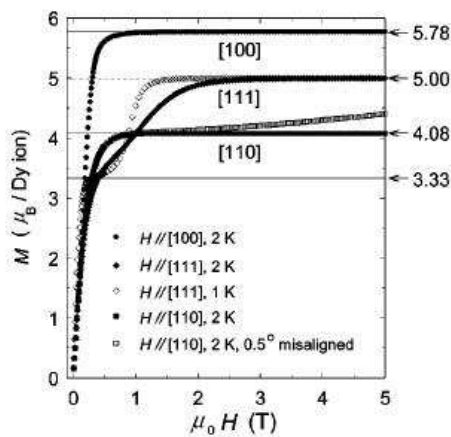}
\end{center}
\vspace{-15pt}
\caption{Magnetization of Dy$_2$Ti$_2$O$_7$ as a function of the applied 
magnetic field  applied along [100], [111], and [110] at low temperatures. 
From H.~Fukazawa {\em et al.} Physical Review B {\bf 65} 054410 (2003). http://link.aps.org/abstract/PRB/v65/p054410   
Copyright 2003 by the American Physical Society.}   
\label{magDy}
\end{figure}

Fig.~\ref{magDy} shows the magnetic field behavior of the magnetization of 
Dy$_2$Ti$_2$O$_7$ at low temperatures for three different directions of the 
field. One can see a good agreement of the theory and experimental 
observations.
 
Other important experimental results for the magnetic field behavior 
of Dy$_2$Ti$_2$O$_7$ and Ho$_2$Ti$_2$O$_7$ the reader can find, e.g., in 
\cite{FBMMW,magn,PetLeeBal,Sak,HiFuMa,Aok,Hig,SaiHigMa,HigMa,Tab,Clan,Mal,
PetLeeBal1,Krey,Matt,Har1}.

\subsection{Dynamics of spin ices} 

The dynamical properties of Dy and Ho pyrochlore oxides in the spin ice phase 
have been intensively studied during the last decade 
\cite{Fuz,SaiHigMa,Matt,Fuk,Sny,Sny1,Sny2,Sny3,Uel,Sut,Kit,Slob,Quil,Klem,
LinLiaChe}.  

\begin{figure}
\begin{center}
\vspace{-15pt}
\includegraphics[scale=0.5]{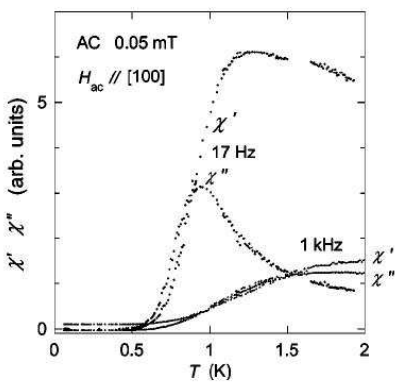}
\end{center}
\vspace{-15pt}
\caption{The temperature dependence of the ac magnetic susceptibility of 
Dy$_2$Ti$_2$O$_7$. From H.~Fukazawa {\em et al.}, Physical Review B {\bf 65}, 
054410 (2002). http://link.aps.org/abstract/PRB/v65/p054410 
Copyright 2002 by the American Physical Society.}   
\label{acsusc}
\end{figure}

\begin{figure}
\begin{center}
\vspace{-15pt}
\includegraphics[scale=0.6]{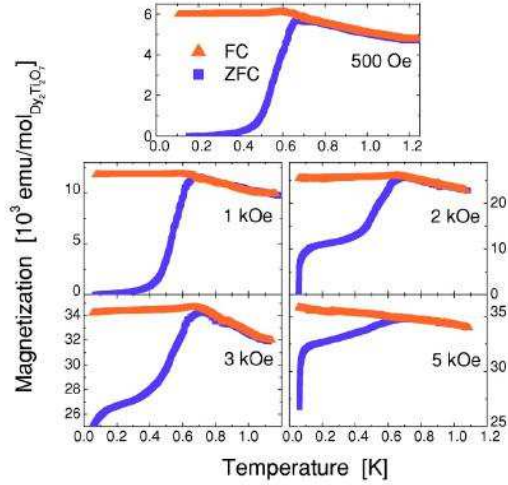}
\end{center}
\vspace{-15pt}
\caption{(Color online) The temperature dependence of the magnetization of 
Dy$_2$Ti$_2$O$_7$ on warming after field cooling (light red curves) and zero 
field cooling (dark blue curves). From J.~Snyder {\em et al.}, Physical Review 
B {\bf 69}, 064414 (2004). http://link.aps.org/abstract/PRB/v69/p064414  
Copyright 2004 by the American Physical Society.}   
\label{zfcfc}
\end{figure}

Low-temperature (down to 0.06~K) low-frequency (about 10~Hz) measurements of 
the ac magnetic susceptibility observed that the real part of the dynamical 
susceptibility becomes lower below about 1~K, while its imaginary part 
manifests a maximum, see Fig.~\ref{acsusc}. Below 0.5~K both real and 
imaginary parts of the dynamical susceptibility have almost zero value, which 
can be explained by absence of a long-range ordering. Such a behavior is 
reminiscent of the behavior of the dynamical characteristics of spin glasses 
\cite{sg,EdAnd,SherKirk,MezParVir,FishHer,Myd,BY}, for which the difference in 
the ``zero field cooling'' and ``field cooling'' behavior is usual (cf. also 
the previous section, where we considered static characteristics of spin ices 
in the external magnetic field). The example of the ``zero field cooled'' and 
``field cooled'' behavior of the magnetization of Dy$_2$Ti$_2$O$_7$ is shown 
in Fig.~\ref{zfcfc}. Similar features in the temperature behavior is seen in 
the real and imaginary part of the dielectric constant, where the external 
magnetic field strongly affects dynamics \cite{SaiHigMa}.

The analysis of the temperature behavior of the real and imaginary parts of the 
dynamical magnetic susceptibility implies the Arrhenius law (while opposite 
conclusions were also made) \cite{Matt,Fuk,Sny,Sny1,Sny2,Sny3,Uel,Sut,Kit,Slob,
Quil,Klem,LinLiaChe}. It turns out that the freezing dynamics of spin ice 
systems differs from the one, associated with spin glasses, where randomness 
plays, probably, the essential role 
\cite{sg,EdAnd,SherKirk,MezParVir,FishHer,Myd,BY}. For example, the behavior of 
dynamical characteristics in the external magnetic field for spin ices is very 
different from the one in spin glasses, where the ordering temperature 
decreases with the growth of the field value. In Dy$_2$Ti$_2$O$_7$ the 
temperature of the feature in the dynamical magnetic susceptibility increases 
with the external field. Also, as expected, the measurements of the dynamical 
characteristics of spin ices have manifested the anisotropy of properties for 
the field applied along [111] and [100]. 

For higher-temperature ($T > 4$~K) dynamical characteristics of spin ices also 
manifest the ``freezing'' feature about 15~K 
\cite{Matt,Fuk,Sny,Sny1,Sny2,Sny3,Uel,Sut,Kit,Slob,Quil,Klem,LinLiaChe}. The 
analysis of that behavior implies the presence of the relaxation process with 
the typical time scales satisfying Arrhenius law $\tau \sim \exp (E_a/T)$ with 
activation energies $E_a \sim 200$~K, which agrees with recent muon spin 
relaxation studies \cite{musr,Que}, which give the muon relaxation rate of 
similar form $\propto \exp(-E_a/T)$ with $E_a \sim 220$~K. It means that such 
a relaxation involves transitions to the higher-energy single-ion multiplets 
of rare earth ions. On the other hand, another dynamical processes in spin 
ices in the temperature range between 5~K and 10~K did not show any significant 
temperature dependence \cite{musr,Que,Ehl}, which has been interpreted as 
caused by the quantum tunneling effect between up- and down-directed states of 
effective Ising spins. Notice that $\mu$SR (muon spin relaxation) and ac 
susceptibility measurements observed relaxation rates different of each other 
up to three orders of magnitude, perhaps, because of the local character of 
the $\mu$SR probe. Depolarization of muons is caused mostly by the development 
on cooling of strong inhomogeneous internal fields (almost static). At very low 
temperatures, inside the spin ice phase, the residual spin dynamics persists 
mostly due to the mixture of electron and nuclear energy levels. Spin dynamics 
in spin ices persists down to lowest temperatures, which fact is observed by 
several experimental techniques \cite{Matt,Fuk,Sny,Sny1,Sny2,Sny3,Uel,Sut,Kit,
Slob,Quil,Klem,LinLiaChe,musr,Que,Erf}. Magneto-caloric studies of 
Dy$_2$Ti$_2$O$_7$ revealed extremely slow relaxation \cite{calor}. We would 
like also to mention here the studies of the dynamics of spin ices via the 
nuclear spin excitation \cite{nucl}, and studies of elastic properties of spin 
ices \cite{Erf,elast}. The latter \cite{Erf} manifested features in the sound 
characteristics of spin ice systems (sound velocity and attenuation) different 
for increasing and decreasing external field value, which also implies 
different relaxation processes in spin ices at low temperatures of order of 
0.03~K. 

Studies of other representatives of the spin ice family, namely, of stannates,  
Dy$_2$Sn$_2$O$_7$ and Ho$_2$Sn$_2$O$_7$ were performed in 
\cite{Sn,Kad,Ehl1,Zhou}.

\section{Monopoles as emergent quasiparticles} 

The interest in spin ices has been considerably grown after the pioneering 
suggestion that magnetic monopoles can exist as emerging quasiparticles there 
\cite{CMS}.  

In physics, the term ``emergence'' is used to describe a phenomenon, which can 
exist at macroscopic scales (in space or time) but not at microscopic scales, 
despite such a macroscopic system can be considered as a large ensemble of 
microscopic systems. It is used to distinguish which laws can be applied to 
macroscopic scales, and which ones only to microscopic scales. Examples of 
emergent macroscopic characteristics can be a temperature in statistical 
mechanics, a convection in liquids or gases. Even a mass, space and time in 
some field theories can be considered as emergent phenomena caused by more 
fundamental concepts, as strings, branes, or Higgs boson. For the recent 
example of emergent phenomenon in frustrated magnetic systems, see, e.g., 
\cite{Lee1}. 

\subsection{Magnetic monopoles}

By magnetic monopoles we usually mean a particle that is an isolated magnet 
with only one magnetic pole. Already in 19th century P.~Curie pointed out that 
magnetic monopoles could exist. However, mainly the problem of magnetic 
monopoles is associated with P.A.M.~Dirac, who has constructed the quantum 
theory of magnetic monopoles \cite{Dir}. He has emphasized that quantum 
mechanics did not preclude the existence of magnetic monopoles, and shown, in 
particular, that if magnetic monopoles existed, then the electric charge had 
to be quantized. We know, naturally, that the electric charge {\em is} 
quantized, however this fact, unfortunately, does not prove the existence of 
magnetic monopoles.   

Standard Maxwell's equations describe magnetism as related to the motion of 
electric charges (take into account that in quantum mechanics particles can 
have ``intrinsic'' magnetic moment related to their spin). The multipole 
expansion produces first monopole, then dipole, quadrupole, etc. For the 
electric field the multipole expansion can have the monopole term (charge), 
while for the magnetic field there is no such a term. That is why, Maxwell's 
equations describe electric charges, but not magnetic charges, despite they 
are symmetric with respect to the interchange of magnetic and electric fields 
(except of the absence of magnetic charges). On the other hand, we can 
formally write symmetric Maxwell's equations, with magnetic monopoles. Two of 
Maxwell's equations, Gauss's law, and Amp\`ere's law, are not changed
(here we use SI units)
\begin{eqnarray} 
&&(\nabla \cdot {\bf E}) = {\rho_e\over \epsilon_0} \ , \nonumber \\
&&[\nabla \times {\bf B}] = \mu_0\epsilon_0 {\partial {\bf E} \over 
\partial t} + \mu_0 {\bf j}_e \ , 
\label{ME1}
\end{eqnarray}
where $\rho_e$ and ${\bf j}_e$ are the electric charge density and electric 
current density, respectively (notice that the vacuum permittivity is 
$\epsilon_0 =c^{-2}\mu_0^{-1} = 8.85\times10^{-12}$~F~m${-1}$, and 
${\bf B}=\mu_0 {\bf H}$). We can re-write Gauss's law for magnetism and 
Faradey's law of induction in such a way that they become similar to 
Eqs.~(\ref{ME1}), namely, 
\begin{eqnarray} 
&&(\nabla \cdot {\bf B}) = \mu_0 \rho_m \ , \nonumber \\
&&-[\nabla \times {\bf E}] = {\partial {\bf B} \over \partial t} 
+ \mu_0{\bf j}_m \ , 
\label{ME2}
\end{eqnarray}
where the magnetic charge density and magnetic current density are introduced. 
Then the Lorentz force can be presented as
\begin{equation}
{\bf F} = q_e({\bf E} + [{\bf v} \times {\bf B}]) +  q_m\left({\bf B} - 
{1\over c^2}[{\bf v} \times {\bf E}]\right) \ , 
\end{equation}
where $q_e$ ($q_m$) are the electric (magnetic) charge of a particle, which 
moves with the velocity ${\bf v}$. For the quantum system, which consists of 
a single stationary electric charge and a single stationary magnetic charge 
(monopole), the electromagnetic field, surrounding them, has the momentum 
density, equal to the Poynting vector 
${\bf G} = (1/\mu_0)[{\bf E}\times {\bf B}]$. It also has the total 
angular momentum, proportional to $\int d{\bf r} [{\bf r}\times {\bf G}]$, 
proportional to $q_eq_m$. The latter is quantized in units 
of $\hbar$ in quantum mechanics. Then Dirac considered a magnetic charge at 
the origin, which generates the magnetic field $q_m/r^2$,  directed in the 
radial direction (analogous to Coulomb's law). The divergence of ${\bf B}$ is 
equal to zero almost everywhere, except for the origin at $r = 0$. We can 
locally define the vector potential such that the curl of the vector potential 
is $[\nabla \times {\bf A}] = {\bf B}$. Such vector potential cannot be defined 
exactly everywhere, because the divergence of the magnetic field is 
proportional to the Dirac delta function at the origin. Dirac defined one 
vector potential on the ``northern hemisphere'' (above the particle), and 
another one for the ``southern hemisphere''. These two vector potentials are 
matched at the ``equator'', and they differ by a gauge transformation. The 
wave function of an electrically-charged particle that moves around the origin
along the ``equator'' is changed by a phase as in the Aharonov–Bohm-Casher 
effect. This phase is proportional to the electric charge $q_e$ of the moving 
particle and to the magnetic charge $q_m$ of the source. The electric charge 
returns to the same point after the total trip around the sphere. The phase of 
its wave function must be unchanged. It implies that the phase added to the 
wave function must be a multiple of $2\pi$. Hence, Dirac's quantum theory means 
quantization of electric and magnetic charges
\begin{equation}
q_eq_m = {2\pi \hbar n\over \mu_0} \ ,  
\end{equation}
where $n$ is an integer. Actually, Dirac's theory describes an infinitesimal 
line solenoid as in the Aharonov-Bohm-Casher effect \cite{AB,AC}, ending at a 
point. The location of the solenoid is the singular part of the Dirac solution. 
This line is known now as Dirac's string. A Dirac string connects monopoles 
and antimonopoles (magnetic particles with opposite to monopole's 
magnetic charge). Dirac strings cannot be seen, because we can put them 
anywhere. For two coordinate patches, we can made the field in each patch 
non-singular by sliding the Dirac string to the place, where we cannot observe 
it. For the recent status of theory and experiment in physics of magnetic 
monopoles consult \cite{Mil}.

\subsection{Magnetic monopoles in spin ices}

In condensed matter, and, in particular, in spin ices, we have, of course, 
$(\nabla \cdot {\bf B})=0$ for the magnetic induction, i.e., no real magnetic 
monopoles should exist. However, there is no such a restriction on the 
microscopically determined magnetic field ${\bf H}$. Hence, we can consider 
quasiparticles, which cannot be constructed as combinations of elementary 
charges; they can carry fractional charges. Namely such quasiparticles can be 
monopoles in the terms of ${\bf H}$.  

\begin{figure}
\begin{center}
\vspace{-15pt}
\includegraphics[scale=0.5]{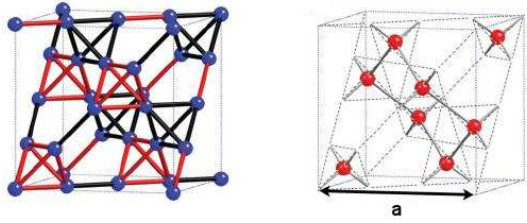}
\end{center}
\vspace{-15pt}
\caption{(Color online) Right panel: Original pyrochlore lattice [A-sublattice 
in light (red); B-sublattice in dark (black)]. Left panel: Dual diamond 
lattice with bonds defining ``easy axes'' for total moments of rare earth ions 
in spin ices. From O.~Benton, O.~Sikora, and N.~Shannon, Physical Review B 
{\bf 86}, 075154 (2012). http://link.aps.org/abstract/PRB/v86/p075154 
Copyright 2012 by the American Physical Society.}   
\label{diamond}
\end{figure}

The ground state of the spin ice can be considered as all tetrahedra obeying 
ice rule (two effective Ising spins are directed ``inside'' and two directed 
``outside'' of each tetrahedron). Effective spins are constrained to be 
directed along their local Ising axes ${\bf e}_n$, which form the diamond 
lattice (dual to the original pyrochlore lattice with vertices at the centers 
of tetrahedra, 
see Fig.~\ref{diamond}) bonds \cite{Ryzh}. To remind, the diamond lattice 
consists of two inter-penetrating FCC sublattices. There is a huge degeneracy 
of such a state, related to the Pauling entropy. Excitations above such a 
ground state manifold are defects, which locally violate the ice rule. Using 
the analogy between the water ice and spin ice \cite{Nag2} we can replace the 
energy of Ising spins living on pyrochlore lattice sites by the energy of 
dipoles (dumbbells consisting of equal in value and opposite in sign magnetic 
charges) that live at the ends of diamond bonds. Let us denote $a_d \equiv 
\sqrt{3/2}a$, which is the diamond lattice constant.  

Let us describe how magnetic monopoles can exist in dipolar spin ices following 
\cite{CMS}. Consider the Hamiltonian ${\cal H}_{ex} + {\cal H}_d$, see 
Eqs.~(\ref{He}) and (\ref{Hd}). A dipole can be thought as a pair of equal and 
opposite charges $\pm q$ separated by the distance ${\tilde a}$, $\mu = 
q{\tilde a}$. Let us choose ${\tilde a} = a_d$, then magnetic charges are 
$q=\mu/a_d$, where $\mu = g\mu_{\rm B}|\langle J_z\rangle |$. The limit 
${\tilde a} \to 0$ reproduces exactly the Hamiltonian (\ref{Hd}). The magnetic 
Coulomb interaction energy between charges situated at different sites of the 
diamond lattice is given by 
\begin{equation}  
v(r_{ij}) = {\mu_0\over 4\pi} {q_iq_j\over r_{ij}}  \ , 
\end{equation}
where $r_{ij}$ is the distance between charges, and we can write such an 
energy for two charges situated at the same site as 
\begin{equation}
v(0) = v_0q_iq_j \ , 
\end{equation}
where we tune the value of $v_0$ to match the interaction energy between two 
neighboring effective Ising spins on the pyrochlore lattice, 
$J_{eff} = \pm (J+5D)/3$ (the latter can be obviously obtained when 
considering one primitive cell). For two neighboring effective spins directed 
inside the tetrahedron, we get
\begin{equation}
J_{eff} = v(0) -2v(r_{12}) -2v(r_{23}) +v(r_{13}) 
\end{equation}
where 1, 2, and 3 define the positions of spins (we have $r_{12} =r_{23}=a_d$ 
and $r_{13}=2a$), while for two spins, one of which is directed ``in'' and the 
other one ``out'' we obtain
\begin{equation}
-J_{eff} = -v(0) +2v(r_{12}) +2v(r_{23}) -v(r_{13}) \ . 
\end{equation}
From these equations we obtain
\begin{equation}
v(0) = J_{eff} +2v(r_{12}) -v(r_{13}) \ , 
\end{equation} 
which yields (using the values for charges $|q_{i,j}| = \mu/a_d$) 
\begin{equation}
v_0 = \left({a_d\over \mu}\right)^2 \left({J\over 3} 
+{4D\over3}\left[1+\sqrt{{2\over3}}\right]\right) \ . 
\end{equation}
Then we can introduce the total magnetic charge on each site $n$ of the 
diamond lattice $Q_n= q_{i1}+q_{i2}+q_{i3}+q_{i4}$ for four charges with the 
coordinates $r_{i,1,2,3,4} =r_{n}$. Then for the Coulomb energy of magnetic 
charges we can write for $n\ne m$ 
\begin{equation}
V(r_{nm}) = {\mu_0\over 4\pi} {Q_nQ_m\over r_{nm} } \ , 
\end{equation}
and for $n=m$ we get $V(r_{nn}) =v_0Q_n^2/2$, which agrees with the above 
values for $v(r)$ up to overall constant term $N(\mu/a_d)^2$, where $N$ is 
the number of dipoles. The energy $v_0/2$ is necessary to reproduce correctly 
the net nearest neighbor interaction. We emphasize that this energy is 
equivalent to the energy of magnetic dipole-dipole interaction between 
effective Ising spins, ${\cal H}_d$. 

\begin{figure}
\begin{center}
\vspace{-15pt}
\includegraphics[scale=0.5]{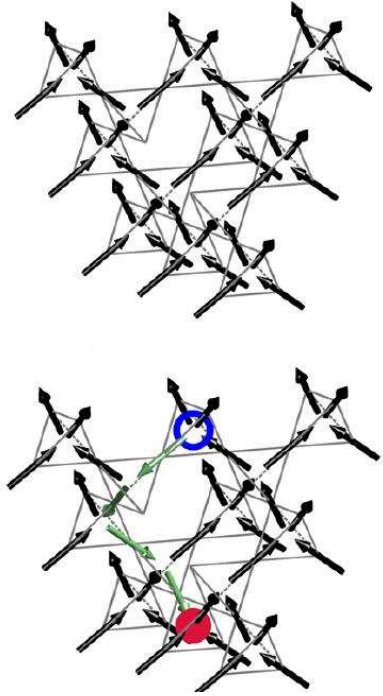}
\end{center}
\vspace{-15pt}
\caption{(Color online) Upper panel: Spin ice, which satisfies the spin ice 
(tetrahedron) rule. Lower Panel: Monopole (blue circle)- antimonopole (red 
circle) pair. Light (green) arrows show possible Dirac's string. From Y.~Wan 
and O.~Tchernyshyov, Physical Review Letters {\bf 108}, 247210 (2012). http://link.aps.org/abstract/PRL/v108/p247210 
Copyright 2009 by the American Physical Society.}   
\label{monopole}
\end{figure}

Let us first consider the ground state of the dipolar spin ice using the 
language of such magnetic charges. The total energy has its minimum if each 
diamond lattice site is neutral, which corresponds to the orientation of 
dipoles such that $Q_n=0$ for each site of the diamond lattice. It is nothing 
else than the realization of the ice (tetrahedron) rule. Naturally, such a 
state is degenerate, which yields the Pauling remnant entropy. Then, let us 
turn to excited states \cite{Ryzh}. Naively the most elementary excitation 
corresponds to the reversing of a single dipole, which generates a local net 
dipole moment $2\mu$. However, such a simple picture is misleading. The 
reversed dipole is related to two adjacent sites with the net magnetic charge 
\begin{equation}
Q_n = \pm {2\mu \over a_d} \ ,  
\end{equation}   
which is the nearest neighbor monopole-antimonopole pair. It is easy to see 
that monopoles can be separated from antimonopoles without violation of the 
ice rule by reversing a chain of adjacent dipoles, or changing the direction of 
effective Ising spins on the original pyrochlore lattice \cite{Ryzh}, see 
Fig.~\ref{monopole}.

A pair of monopoles separated by the distance $r$ experience a Coulomb magnetic 
coupling $-\mu_0 q_m^2/4\pi r$. It takes only a {\em finite} energy to 
separate monopoles to infinity, which means that monopoles are deconfined. 
Hence, magnetic monopoles are are true elementary excitations of the spin ice. 
They are emergent quasiparticles, because of the fractionalization of their 
charge, see above. The fact that a string of dipoles realizes a 
monopole-antimonopole pair at its ends is know from the classical 
electrodynamics \cite{Jack}. However, it is important that the energy cost of 
creating such a string of dipoles remains bounded with the growth of its length
(the relevant string tension vanishes) to obtain deconfined monopoles. Of 
course, such a condition cannot be realized in a vacuum, where by growing the 
length of the string of dipoles we need the energy for creation of additional 
dipoles. The ice rule can be considered as the requirement that two dipole 
strings enter and exit each site of the diamond lattice. No domain walls are 
created along the string in the dipolar spin ice (unlike, e.g, an ordered 
ferromagnet), which results in the deconfinement of monopoles there.   

According to the Dirac quantization condition, the charge of real magnetic 
monopoles has to be quantized, which is in the close relation to the condition 
that Dirac's string is unobservable. On the other hand, the net of (dipole) 
strings in the dipolar spin ice, which are energetically unimportant, makes 
dipole strings in such spin ices observable, and the magnetic charge there is 
not quantized. We can define a density of ``smeared'' magnetic charges in the 
dipolar spin ice as 
\begin{equation}
\rho_m({\bf r}) = \int d^3{\bf r}' (\nabla \cdot {\bf H})
e^{-|{\bf r}-{\bf r}'|^2/\xi^2} \ , 
\end{equation} 
where the monopole at the origin ${\bf r}=0$ separated by $L \gg \xi \gg a$ 
from other monopoles yields $\rho_m(0) =\pm q_m$. For the magnetic induction 
${\bf B}$, the compensating flux moves along ``non-quantized Dirac's 
string'' of flipped dipoles, created together with each monopoles. 

\subsection{Properties of magnetic monopoles in spin ices} 

The external magnetic field applied along [111] acts as a staggered chemical 
potential for monopoles \cite{CMS}. We can approximate the low-energy physics 
of dipolar spin ices as the one of the gas of magnetic monopoles and 
antimonopoles on the diamond lattice. Hence, one can use the results for such 
a gas with Coulomb coupling \cite{Fish}. By changing the value of the chemical 
potential in that case one can see the temperature crossover between high- and 
low-density phases at high temperatures, while at low temperatures there must 
be the first order phase transition between those phases. The line of that 
phase transition terminates in a critical point in the phase diagram. Notice 
that for nearest-neighbor spin ice such a liquid-gas transition cannot exist 
\cite{Li2}, there defects interact only entropically. 
The low-density phase of the gas of monopoles is related to the Kagome phase in 
[111] magnetic field, while the high-density phase corresponds to the ordered 
state with the maximal magnetization along the field direction. Notice that 
monopoles, which appear in the anomalous Hall effect \cite{Fang}, are not 
excitations, and involve a real physical magnetic field.  

Let us now calculate the equilibrium concentration of monopoles \cite{Ryzh}. 
Each vertex in the diamond lattice at low energies can be in one of 14 states: 
six monopole-free states (satisfying the ice rule), four states with a 
monopole, and four states with an antimonopole (three effective Ising spins 
directed ``in'' and one ``out'', and vice versa). Two states with all 
effective spins directed ``inside'' or ``outside'' the tetrahedron can be 
ignored in the low energy theory. If $N_t$ is the number of tetrahedra, and 
the number of vertices for each of such states is $N_i$ ($i=1,\dots,14$), then 
the total number of configurations is $N_t\!/\prod_i N_i$ The configurations 
with parallel and antiparallel effective spins at the midpoints of nearest 
neighbor bonds (in other words, with correlated and anti-correlated vertices) 
must not be counted. If the probability of a correlated state is 1/2 then the 
number of correlated states is $w_{\pm} =(1/2)^{2N_t}N_t\!/\prod_iN_i$. If 
monopoles and antimonopoles are created in pairs, and all states with 
monopoles and antimonopoles are equivalent $N_1 = \dots = N_6$, 
$N_7 =\dots = N_{14}$, the entropy per rare earth ion is 
$-k_{\rm B}(2x \ln (x) +(1-2x)\ln [2(1-2x)/3])$ where $x=N_{\pm}/N_t$ is the 
concentration of monopoles (antimonopoles) per vertex. The free energy per 
vertex is then ($\varepsilon_{\pm}$ are the energies of the 
monopole/antimonopole configurations) 
\begin{eqnarray}
&&f = \varepsilon_{\pm} x +k_{\rm B} T \biggl[ 2x\ln (x) 
\nonumber \\   
&&+(1-2x)\ln\left({2(1-2x)\over 3}\right)\biggr] \ ,  
\end{eqnarray}
which implies the equilibrium concentration of monopoles being 
\begin{equation}
x_{\pm} = {2\exp(-\varepsilon_{\pm}/2k_{\rm B}T)\over 
3+ 4\exp(-\varepsilon_{\pm}/2k_{\rm B}T)} \ . 
\end {equation}
At low temperatures, which is relevant for the experimental situation in spin 
ice materials, we have $x_{\pm} \approx (2/3)
\exp(-\varepsilon_{\pm} /2k_{\rm B}T)$. On the other hand, at high enough 
temperatures we get, obviously, $x_{\pm} =2/7$.  

The monopole picture of spin ices is different from the conventional one in the 
theory of quasiparticles, because the ground state, as well as excited states 
for monopoles are highly degenerate. However, for many purposes the information 
about local configurations is redundant. 

Using the analogy between the system of monopoles in the spin ice and water 
ice it is possible to write the `continuity' equation for the magnetization 
${\bf M}$
\begin{equation}
{\partial {\bf M} \over \partial t} = {Q \over V} (N_+{\bf v}_+ -N_-{\bf v}_-) 
\ , 
\end{equation} 
where $Q \equiv Q_n$ is the monopole magnetic charge, $V$ is the 
macroscopically small volume around the point ${\bf r}$, and ${\bf v}_{\pm}$ 
are the velocities of the monopole and antimonopole. Notice that ${\bf j}_{\pm} 
={\bf v}_{\pm}N_{\pm}/V$ are densities of currents for monopoles 
(antimonopoles). The rate of the entropy production due to monopoles 
(antimonopoles) can be written as
\begin{equation} 
T{\partial {\cal S}\over \partial t} = \sum_{\pm} \biggl[\pm {\bf j}_{\pm} 
\biggl(Q{\bf H} - {8ak_{\rm B}T{\bf M}\over \sqrt{3}Q} \biggr)\biggr] \ .  
\end{equation}
On the other hand, it is equal to $\sum_{\pm} {\bf j}_{\pm} {\bf f}_{\pm}$, 
where ${\bf f}_{\pm}$ are generalized driving forces. It follows that 
${\bf f}_{\pm} = \pm (Q{\bf H} \mp 8ak_{\rm B}T{\bf M}/\sqrt{3}Q)$. The second 
term describes possible magnetic ordering: When effective spin are partly 
ordered there exists a nonzero monopole current even without the external 
field. Then the monopole (antimonopole) currents can be written as 
\begin{equation}
{\bf j}_{\pm} = \pm u_{\pm} x_{\pm}N_t\left(Q{\bf H} -
{8ak_{\rm B}T{\bf M}\over \sqrt{3}Q}\right) \ , 
\end{equation}
where $u_{\pm}$ are the monopole/antimonopole mobilities, so that 
$\kappa_{\pm} = u_{\pm} x_{\pm}Q$ are the monopole/antimonopole conductivities.  
Then it is easy to obtain, taking into account the `continuity equation', that 
in the linear regime ${\bf M}_{\omega} = \chi(\omega) {\bf H}_{\omega}$ with the 
longitudinal dynamical magnetic susceptibility
\begin{equation} 
 \chi(\omega) = {(\sqrt{3} Q^2 /8a k_{\rm B} T)\over 1-i\omega \tau} \ , 
\end{equation}
where $\tau$ is the relaxation time. The static magnetic susceptibility 
can be then obtained as 
\begin{equation}
\chi_T = {\sqrt{3} g^2\mu_{\rm B}^2|\langle J_z\rangle |^2 \over a^3 k_{\rm B} T} 
\ , 
\end{equation}
The absolute value of the static susceptibility is twice the value of the 
susceptibility of the standard paramagnet $\chi_C$ of the same spin density. 
This expression for the homogeneous susceptibility has been generalized 
recently for the inhomogeneous case \cite{Bram1}, when taking into account 
the diffusion of monopoles, as
\begin{equation}
\chi({\bf q},\omega) = {\chi_T\over 1+(a^2q^2/6gx) -i\omega \tau} \ , 
\end{equation}
where ${\bf q}$ is the wave vector, $x=8a^3c/3\sqrt{3}$ is the total 
dimensionless monopole density ($c$ is the total concentration of monopoles) 
and $g=\chi_C/\chi_T$ is the ratio of the static susceptibilities for the spin 
ice and standard paramagnet; it is equal to 1/2 in the above calculations, 
however, in general, it can vary from 1 at high temperatures to 1/2 at low 
temperatures. It implies the correlation length $\xi = a/\sqrt{6gx}$. 

If we take into account the demagnetization factor ${\cal D}$ via 
${\bf H}_{int} ={\bf H}_{ext} -{\cal D} {\bf M}$, the effective susceptibility 
has to be renormalized as $\chi_r \equiv {\bf M}/{\bf H}_{ext} = \chi_T /
({\cal D} \chi_T +1)$. Then the ``field cooled'' magnetization is just 
${\bf M}_{fc} = \chi_r {\bf H}_{ext}$, however, the ``zero field cooled'' one 
is ${\bf M}_{zfc} = \chi_r {\bf H}_{ext}[1-\exp(-t/\tau)]$ (valid at small 
$t \ll \kappa_{\pm}$), where the time-dependent multiplier comes from the 
integration of the `continuity equation' for magnetization when we take into 
account relaxation time $\tau$ and demagnetization factor ${\cal D}$. The 
behavior of the magnetization derived from that theory \cite{Bram1} is 
reminiscent of the experimentally observed in spin ices data, presented in 
Fig.~\ref{zfcfc}.  

Closely related problem of magnetic relaxation in spin ices as a ``monopole 
electrolyte'' has been studied in \cite{MoShif,JH,JH1}. Non-Ohmic 
conductivity, the Wien effect, \cite{Ons} for a weak ``monopole electrolyte'' 
has been studied theoretically and experimentally by the transverse field 
low-temperature $\mu$SR \cite{Wien,Gibl} (notice, though \cite{Dun}). Other 
recent theoretical studies of the dynamical characteristics of monopoles 
include \cite{Bram,StyFei,RyzRyz,RyzKlu,Blu,TakTat,Stre}.  

Some low-temperature properties of spin ices were successfully described by 
magnetic monopoles, e.g., in neutron scattering \cite{Morr,FenDee}, in the 
behavior of magnetic susceptibility \cite{Kadow} (the latter can 
be well described by the Debye-H\"uckel theory \cite{CMS3,Zhou1,CMS2}), see 
also \cite{Matt,other,Yar,Rev}, in NMR (nuclear magnetic resonance) 
\cite{Sala}, and in the thermal conductivity \cite{Koll}.   

\section{Other spin ices} 

So far we discussed properties of the standard spin ices. However, nowadays 
physicists consider other possibilities for spin ices. 

\subsection{Quantum spin ice}

\begin{figure}
\begin{center}
\vspace{-15pt}
\includegraphics[scale=0.5]{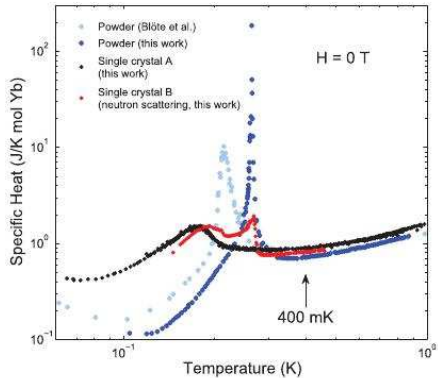}
\end{center}
\vspace{-15pt}
\caption{(Color online) The temperature dependence of the specific heat of 
Yb$_2$Ti$_2$O$_7$ for powder samples (dark and light blue curves) and 
single crystals (red and black curves). From K.A.~Ross {\em et al.}, Physical 
Review B {\bf 84}, 174442 (2011). http://link.aps.org/abstract/PRB/v84/p174442 
Copyright 2011 by the American Physical Society.}   
\label{qsi}
\end{figure}

By the quantum spin ice we usually mean similar rare earth pyrochlore oxides, 
in which, unlike usual spin ices, the ``easy axes'' magnetic anisotropy is not 
so strong comparing to the spin-spin (total moment-total moment) interaction, 
like the exchange coupling, or the magnetic dipole-dipole interaction. That is 
why, there exists a possibility of spreading of a local spin flip to other 
places due to non-Ising components of the particle-particle interaction 
\cite{Moe,qsi,MoeTcheSon,Bur,MaeOySa,OnTa,Ross,Shan,WanTch,BenSikSha,LeeOnBal}. 
In general, such a possibility can yield magnetic ordering, hence, the level 
of magnetic frustration for quantum spin ices is lower than for usual ones 
(sometimes called classical). Strong quantum fluctuations essentially 
affect statics and dynamics of quantum spin ices. Recent studies show that 
Yb-based titanate, Yb$_2$Ti$_2$O$_7$, can serve as a good example of a quantum 
spin ice \cite{Yb,Hod,Ross1,Yas,Gar,Cao,Yao,Ross2,Tom,Tom1,Ross3,Lho,Appl}. 
Crystalline electric field also well separates the low-energy doublet from 
other multiplets there. However, planar components of $g$-tensor, 
$g_{\perp} =4.18$ are larger than longitudinal Ising components (along [111]) 
$g_{\parallel} =1.77$. Notice that in this compound the anisotropy is also 
present in exchange interactions with the ferromagnetic $\theta_{\rm CW} 
\approx 0.65 \pm 0.15$~K. The system reveals the phase transition at 0.24~K to 
the low-temperature phase (the value of the critical temperature depends on 
the applied magnetic field), which nature has not been totally identified yet, 
see, e.g. Fig.~\ref{qsi}. 

Most of studies support ferromagnetic ordering in that compound. This phase 
transition has been recently determined as the Higgs transition from a 
magnetic Coulomb liquid of monopoles to the ferromagnetic phase, which was 
viewed as a Higgs phase for magnetic monopoles \cite{Pow,Chang}.

\subsection{Stuffed spin ice} 

By stuffed spin ice \cite{GGG} one means the situation, when magnetic 
rare-earth ions alter chemically non-magnetic Ti sites, for example 
Ho$^{3+}$ ``stuffs'' B-sites like in Ho$_2$(Ti$_{2-x}$Ho$_x$)O$_{7-\delta}$ (where 
$\delta >0$ implies the balance of oxygen content due to ``stuffing'') 
\cite{Hostuf,Lau,Lau1,Zhou2,Lau2,Ehl2,Gar1,Ald}. Such a procedure, naturally, 
introduces randomness to the spin ice, and one would expect enhancement of 
spin glass-like behavior, e.g., the transition to the ordered spin-glass 
state. The quantum relaxation time is enhanced there, i.e., spin-spin 
correlations are slower in stuffed spin ices. However, stuffed spin ices do 
not freeze down to the lowest temperatures, and have basically the same 
entropy as standard spin ices \cite{Hostuf,Lau,Lau1,Zhou2,Lau2,Ehl2,Gar1,Ald}. 
On the other hand, stuffed spin ice based on Dy does not manifest residual 
entropy, i.e., spin fluctuations persist there down to lowest temperatures 
\cite{Rev,Dystuf}. Extra spins of Dy-based stuffed spin ice trap magnetic 
monopoles and obstruct flow of monopoles, introducing residual resistance.
For Ho-based stuffed spin ice the ice rules are valid only over a short range. 
At longer range such a stuffed spin ice exhibits some characteristics of a 
``cluster glass'', with a tendency to more conventional ferromagnetic 
correlations.

\subsection{Metallic spin ice}

We considered above insulating spin ice systems, where the movement of 
electric charges was absent. Some rare-earth pyrochlore oxides, on the other 
hand, reveal conducting properties. For example, Pr$_2$Ti$_2$O$_7$ manifests 
Kondo-like effects (like logarithmic increase of the resistivity and magnetic 
susceptibility at low temperature) \cite{met,Nak,Mach}. It is strange, because 
Pr$^{3+}$ is the Ising ion, and the Ising anisotropy reduces the Kondo 
screening. Theoretical studies predict that long-range RKKY interaction in 
metallic pyrochlore oxides should yield magnetic ordering; on the other hand, 
local spin correlations of the spin ice can produce non-Kondo mechanism of 
observed features in the temperature dependences 
\cite{metteor,UdaIshMot,IshUdMot}. The other metallic spin-liquid system, 
Pr$_2$Ir$_2$O$_7$ with Ising-like spins along [111] reveals spontaneous Hall 
effect \cite{Nak,sH,Mill,Mach1}. There spin-ice correlations in the liquid 
phase lead to a non-coplanar spin texture forming a uniform but hidden order 
parameter, the spin chirality. 

\section{Artificial spin ice} 

\begin{figure}
\begin{center}
\vspace{-15pt}
\includegraphics[scale=0.4]{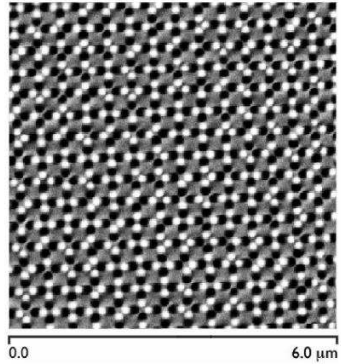}
\end{center}
\vspace{-15pt}
\caption{Picture from the magnetic force microscope of the square array of 
single-domain permalloy magnetic nanoislands. White and black sides of each 
island show the direction of the magnetic moment of the island. From 
C.~Nisoli {\em et al.}, Phys. Rev. Lett. {\bf 98}, 217203 (2007). http://link.aps.org/abstract/PRL/v98/p217203 
Copyright 2009 by the American Physical Society.}   
\label{artif}
\end{figure}

Spin ice is a very interesting system, which, as we have shown above, 
manifests very rich physics. This is why the study of spin ices has not been 
limited by natural systems exhibiting spin ice properties. Several years ago 
the modern lithographic technique was used for the construction of artificial 
dipolar arrays of single-domain ferromagnetic permalloy Ni$_{0.81}$Fe$_{0.19}$ 
islands of a submicron size (with the length of 220~nm, width of 80~nm and 
thickness of 25~nm) on a Si substrates with a native oxide layer \cite{Wa}. 
The moment of each island was about 3$\times 10^{7}\mu_{\rm B}$. Notice that 
permalloy has effectively zero magnetic anisotropy, so that the anisotropy 
energy of the island's magnetic moment itself, which is controlled by it's 
shape (of order of 10$^4$~K), forced magnetic moments of islands to align 
along the longer axes, therefore magnetic islands can be considered as 
effective Ising-like spins. Such arrays of interacting monodomain nanomagnets 
provide important model systems of statistical mechanics, as they map onto 
well studied theoretically vertex models, see above. The intrinsic frustration 
on such a lattice is similar to spin ices. To see how it comes about, we can 
consider a vortex, where four islands meet. A pair of moments in the vortex 
can be directed either to maximize or to minimize the magnetic dipole-dipole 
interaction. It is energetically favorable if the moments of pair of islands 
are directed in a such a way that one is pointing into the center of the 
vortex, and the other is directed out of the center of the vortex. On the 
other hand, the configuration with both moments pointing inside vortex (or 
outside it) demands additional energy. There are in general 16 configurations 
of vortices. Six configurations, satisfying the ice rule,  with ``two-in'' 
and ``two out'', like 5 and 6 of Fig.~\ref{6vert} have the lowest energy (the 
configurations 1, 2, 3 and 4 have higher energies than 5 and 6). On the other 
hand, the configurations with three moments inside (outside) and four moments 
inside (outside) the vertex are energetically unfavorable at low temperatures. 
The energy difference between configurations can be, in principle, regulated by 
the size of the lattice constant. This approach, in particular, gives the 
possibility to study the state of the system with local probes, as the 
magnetic force microscope, to see directly the situation with single 
constituent magnetic islands, see the example in Fig.~\ref{artif}. Magnetic 
monodomain permalloy islands that way provide the analogue of effective Ising 
spins in pyrochlore oxides. 

The construction of artificial frustrated magnet has opened the door to the 
new approach for the researches with designed artificial systems rather than 
with natural ones. For example, artificial spin ice systems were proposed to 
be constructed as arrays of optical traps \cite{optic,Stru}. A large number of 
experiments and theories since 2006 have considered artificial spin ice 
systems, including planes of ferromagnetic islands with square, honeycomb and 
Kagome lattices \cite{asi,LePo,QiBrCu,Meng,Tra,Lib,LiKe,BudPolSta,Shu,Nic,
Tche,Mell,LiWang,LiZh,Rou,LiWa,Dau,Budr1,Rei,Kap,Scu,Budr2,Budr3,Mol,ZhaLi,
Arn,Morg}. In particular, it has been shown using magneto-optical Kerr effect 
that disorder in the roughness (in shape) of magnetic islands plays essential 
role in the collective behavior of artificial spin ices \cite{dis,Budr4}. The 
interesting study has investigated the behavior of entropy in artificial the 
spin ice system \cite{Lamm}. The analysis shows that nearest-neighbor 
correlations drive the longer-range ones there. 

As a result of the magnetic frustration, these systems can exhibit magnetic 
monopole type states, which are an example of an exotic emergent 
quasiparticle \cite{Mell,monart,Mol1,MolMo,RosFra,Morg1,Lad,Pha,Lad1,MorSteLan,
Sil,Hug,Pol,Sil1}. For example, magnetic monopoles and associated 
Dirac-like strings have been directly observed in the artificial honeycomb 
(on Co films of 20~nm thickness) and Kagome spin ice (permalloy films) 
systems \cite{Meng1,Meng2,Cum} using magnetic force microscopy and X-ray 
photoemission microscopy. To remind, the Kagome lattice can be realized in 
pyrochlore spin ices by applying [111] external magnetic field. In particular, 
for the visualization of magnetic monopoles in permalloy systems the X-ray 
magnetic circular dichroism was used. Dirac-like strings were observed as a 
history of propagating monopole-antimonopole pairs. Creation of such pairs as 
well as their movement has been regulated by the external magnetic field in the 
reversed (with respect to the magnetic moments of islands) direction. 
Randomness, as for other artificial spin ices, see above, plays an important 
role for the physics of monopoles. In contrast to pyrochlore oxides, where 
magnetic monopoles form a gas and Dirac's strings are dynamically fluctuating, 
in an artificial spin ice one deals at large enough values of the external 
magnetic field with the effective low-temperature case, in which after each 
field step (change of the magnetization of an island) random variations in the 
switching field pin monopoles and related Dirac's strings. Monopoles become 
trapped. Namely that property permits to image monopole-antimonopole 
configurations before increasing the value of the field, and to manipulate 
with such magnetic charges. Dirac's strings grow in the horizontal or diagonal 
directions of the two-dimensional lattice as a result of one-dimensional 
avalanche processes \cite{aval}. 

Artificial spin ices reveal also the anomalous Hall effect \cite{artaHe}, 
which, like in the ferromagnetic SrRuO$_3$ is believed to be caused by the 
movement of magnetic monopoles \cite{Fang}.  

\section{Summary}

Studies of magnetic frustrated systems nowadays belong to the one of the 
most developing branches of the low-temperature condensed matter physics. It 
is determined by the great variety of new physical concepts, which were 
applied, and plenty of new physical effects, observed in this field. Spin 
ices, magnetic monopoles, Higgs effect, anomalous Kondo and Hall physics: All 
of them have been observed and explained during recent years in frustrated 
magnets. The studies of magnetic frustrated systems are far from being 
complete; many new important and interesting effects are waiting for their 
discoveries. Least but not the last: Frustrated magnets are important not only 
due to their fundamentally interesting physical properties, but also because of 
their perspective usefulness as data storages and memories for computers, or 
as possible realizations of the topological quantum computation. 

It is possible that, when reviewing such a swiftly developing field of physics 
with a great number of important works, and trying to mention all of them, I, 
perhaps, have not cited some interesting publications. I sincere apologize to 
those of authors, whose contributions to the field of spin ices and magnetic 
monopoles there are not mentioned in my article.  

I thank R.~Moessner for his very helpful comments and suggestions. Support 
from the Institute for Chemistry of V.N.~Karazin Kharkov National University 
is acknowledged. 

Preprinted figures with permission from H.~Fukuzawa {\em et al.}, 
Physical Review B {\bf 65}, 054410 (2002); J.~Snyder {\em et al.}, 
Physical Review B {\bf 69}, 064414 (2004); J.S.~Gardner {\em et al.}, 
Physical Review B {\bf 70}, 180404(R) (2004); X.~Ke {\em et al.}, Physical 
Review Letters {\bf 99}, 137203 (2007); C.~Nisoli {\em et al.}, Physical 
Review Letters {\bf 98}, 217203 (2007); J.P.~Clancy {\em et al.}, Physical 
Review B {\bf 79}, 014408 (2009); K.A.~Ross {\em et al.}, Physical Review 
Letters {\bf 103}, 227202 (2009); A.~Yaouanc {\em et al.}, Physical Review B 
{\bf 84}, 172408 (2011); K.A.~Ross {\em et al.}, Physical Review B {\bf 84}, 
174442 (2011); Y.~Wan and O.~Tchernyshyov, Physical Review Letters {\bf 108}, 
247210 (2012); O.~Benton, O.~Sikora, and N.~Shannon, Physical Review B 
{\bf 86}, 075154 (2012), Copyrights 2002, 2004, 2007, 2009, 2011, 2012 by 
the American Physical Society. Readers may view, browse, and/or download 
material for temporary copying purposes only, provided these uses are for 
noncommercial personal purposes. Except as provided by law, this material may 
not be further reproduced, distributed, transmitted, modified, adapted, 
performed, displayed, published, or sold in whole or part, without prior 
written permission from the American Physical Society.


\begin{thebibliography}{999}
\bibitem{Mat} See, e.g., D.C.~Mattis, {\em The Theory of Magnetism I}, 
Springer, Berlin (1988).
\bibitem{Zb} See, for example, A.A. Zvyagin, {\em Quantum Theory of 
One-Dimensional Spin Systems}, Cambridge Scientific Publishers, Cambridge 
(2010). 
\bibitem{Ball} J.C.~Ballhausen, {\em Introduction to Ligand Field Theory}, 
McGraw-Hill, NY (1962). 
\bibitem{vV} J.H.~van Vleck, {\em The Theory of Electric and 
Magnetic Susceptibilities}, Oxford University Press, Oxford (1952). 
\bibitem{AB1} A.~Abragam and B.~Bleaney, {\em Electron Paramagnetic Resonance 
of Transition Ions}, Clarendon, Oxford (1969).   
\bibitem{Stev} K.W.H.~Stevens, Proc. Phys. Soc. A {\bf 65}, 209 (1952). 
\bibitem{Heis} W.~Heisenberg, Z. Phys. {\bf 49}, 619 (1928); 
\bibitem{Dir1} P.A.M.~Dirac, Proc. Roy. Soc. (London) {\bf A123}, 714 (1929). 
\bibitem{RK} M.A.~Ruderman and C.~Kittel, Phys. Rev. {\bf 96}, 99 (1954).
\bibitem{Kas} T.~Kasuya, Progr. Theor. Phys. {\bf 16}, 45 (1956). 
\bibitem{Yos} K.~Yosida, Phys. Rev. {\bf 106}, 893 (1957). 
\bibitem{Smart} See, e.g., J.S.~Smart, {\em Effective Field Theories of 
Magnetism}, W.B.~Saunders Company, Phyladelphia, (1966). 
\bibitem{spinwave} F.~Bloch, Z. Phys. {\bf 61}, 206 (1930). 
\bibitem{HolPrim} T.~Holstein and H.~Primakoff, Phys. Rev. {\bf 58}, 
1098 (1940). 
\bibitem{Neel} L.~N\'eel, Ann. de Phys. {\bf 17}, 64 (1932); 
\bibitem{Neel1} L.~N\"eel, Ann. de Phys. {\bf 18},  5 (1932). 
\bibitem{Con} J.H.~Conway and N.J.A.~Sloane, {\em Sphere Packings, Lattices 
and Groups}, Grundlehren der Mathematischen Wissenschaften {\bf 290}, 
Springer-Verlag, Berlin (1999). 
\bibitem{And1} P.W.~Anderson, Mater. Sci. Bull. {\bf 8}, 1008 (1956); Science, 
{\bf 235}, 1196 (1987). 
\bibitem{Bal1} L.~Balents, Nature {\bf 464}, 199 (2010). 
\bibitem{Toul} J.~Vannimenus and G.~Toulouse, J. Phys. C: Solid St. Phys. 
{\bf 10}, L537 (1977). 
\bibitem{Vill} J.~Villain, J. Phys. C: Solid St. Phys. {\bf 10}, 1717 (1977).  
\bibitem{sg} V.~Canella and J.A.~Mydosh, Phys. Rev. B {\bf 6}, 4220 (1972). 
\bibitem{EdAnd} S.F.~Edwards and P.W.~Anderson, J. Phys. F: Met. Phys. {\bf 5}, 
965 (1975). 
\bibitem{SherKirk} D.~Sherington and S.~Kirkpatrick, Phys. Rev. Lett. 
{\bf 35}, 1792 (1975). 
\bibitem{MezParVir} M.~Mezard, G.~Parisi, and M.A.~Virasoro, {\em Spin glass 
theory and beyond}, World Scientific, Singapore (1987). 
\bibitem{FishHer} K.H.~Fischer and J.A.~Hertz, {\em Spin Glasses}, 
Cambridge University Press, Cambridge (1991). 
\bibitem{Myd} J.A.~Mydosh, {\em Spin Glasses}, Taylor \& Frensis, NY (1995).
\bibitem{Hou} R.M.F.~Houtappel, Physica {\bf 16}, 425 (1950). 
\bibitem{Wan} G.H.~Wannier, Phys. Rev. {\bf 79}, 357 (1950). 
\bibitem{YafKit} Y.~Yafet and C.~Kittel, Phys. Rev. {\bf 87}, 290 (1952).
\bibitem{Diep} {\em Frustrated Spin Systems}, Ed.: H.T.~Diep, World 
Scientific, Singapore (2004).
\bibitem{Lacr} {\em Highly Frustrated Magnetism}, Eds.: C.~Lacroix, 
P.~Mendels, and F.~Mila, Springer, Berlin (2010). 
\bibitem{Ram} A.P.~Ramirez, Annu. Rev. Mater. Sci. {\bf 24}, 453 (1994).
\bibitem{Ram1} A.P.~Ramirez, Annu. Rev. Mater. Sci. {\bf 24}, 453 (1994). 
\bibitem{Gaul} B.D.~Gaulin, Hyperfine Interact. {\bf 85}, 159 (1994). 
\bibitem{ShifRam} P.~Schiffer and A.P.~Ramirez, Comments Condens. Matter. 
Phys. {\bf 18}, 21 (1996). 
\bibitem{Geed} J.E.~Greedan, J. Matt. Chem. {\bf 11}, 37 (2001). 
\bibitem{MoeRam} R.~Moessner and A.R.~Ramirez, Physics Today {\bf 59}, 24 
(2006). 
\bibitem{CMS1} C.~Castelnovo, R.~Moessner, and S.L.~Sondhi, Ann. Rev. Condens. 
Matter Phys. {\bf 3}, 35 (2012).   
\bibitem{strXI} A.J.~Leadbetter, R.C.~Ward, J.W.~Clark, P.A.~Tucker, T.~Matsuo, 
and S.~Suga, J. Chem. Phys. {\bf 82}, 424 (1985). 
\bibitem{Mur} B.~J.~Murray and A.K.~Bertram, Phys.Chem.Chem.Phys. {\bf 8}, 186 
(2006).
\bibitem{fer} M.J.~Iedema, M.J.~Dressler, D.L.~Doering, J.B.~Rowland, 
W.P.~Hess, A.A.~Tsekouras, and J.P.~Cowin, J. Phys. Chem. B {\bf 102}, 9203 
(1998). 
\bibitem{hbond} T.S.~Moore and T.F.~Winmill, J.Chem. Soc., Trans. {\bf 101}, 
1635 (1912). 
\bibitem{hbond1} W.M.~Latimer and W.H.~Rodebush, J. Am. Chem. Soc. {\bf 42}, 
1419 (1920).
\bibitem{BF} J.D.~Bernal and R.H.~Fowler, J. Chem. Phys. {\bf 1}, 515 (1933). 
\bibitem{Gia} W.F.~Giaque and M.F.~Ashley, Phys. Rev. {\bf 43}, 81 (1933).
\bibitem{Gia1} W.F.~Giaque and J.W.~Stout, J. Am. Chem Soc. {\bf 58}, 1144 
(1936). 
\bibitem{Paul} L.~Pauling, J. Am. Chem. Soc. {\bf 57}, 2680 (1935). 
\bibitem{Nag} J.F.~Nagle, J. Math. Phys. {\bf 7}, 1484 (1966). 
\bibitem{Rys} F.~Rys, Helv. Phys. Acta {\bf 36}, 537 (1963). 
\bibitem{Lieb} E.H.~Lieb, Phys Rev. {\bf 162}, 162 (1967). 
\bibitem{Lieb1} E.H.~Lieb, Phys. Rev. Lett. {\bf 18}, 692 (1967). 
\bibitem{Lieb2} E.H.~Lieb, Phys. Rev. Lett. {\bf 19}, 108 (1967). 
\bibitem{Suth} B.~Sutherland, Phys. Rev. Lett. {\bf 19}, 103 (1967).
\bibitem{Y} C.P.~Yang, Phys. Rev. Lett. {\bf 19}, 586 (1967). 
\bibitem{N} J.F.~Nagle, Comm. Math. Phys. {\bf 13}, 62 (1969). 
\bibitem{Bax} R.J.~Baxter, {\em Exactly solved models in statistical 
mechanics}, Academic Press, London (1982). 
\bibitem{Ver} E.J.W.~Verwey, Nature {\bf 144}, 327 (1939). 
\bibitem{VerHaa} E.J.W.~Verwey and P.W.~Haaymann, Physica {\bf 8}, 979 (1941). 
\bibitem{And} P.W.~Anderson, Phys. Rev. {\bf 102}, 1008 (1956). 
\bibitem{Li} S.~Kondo, D.C.~Johnston, C.A.~Swenson, F.~Borsa, A.V.~Mahajan, 
L.L.~Miller, T.~Gu, A.I.~Goldman, M.B.~Maple, D.A.~Gajewski, E.J.~Treemann, 
N.R.~Dilley, R.P.~Dickey, J.~Merrin, K.~Kojima, G.M.~Luke,Y,J.~Uemura, 
O.~Chmaissem, and J.D.~Jorgensen, Phys. Rev. Lett. {\bf 78}, 3729 (1997). 
\bibitem{KonJonMil} S.~Kondo, D.C.~Johnston, and L.L.~Miller, Phys. Rev. B 
{\bf 59}, 2609 (1999).
\bibitem{Chmai} O.~Chmaissem, J.D.~Jorgensen, S.~Kondo, and D.C.~Johnston, 
Phys. Rev. Lett. {\bf 79}, 4866 (1997). 
\bibitem{Ey} V.~Eyert, K.H.~Hock, S.~Horn, A.~Loidl, and P.S.~Resiborough, 
Europhys. Lett {\bf 46}, 762 (1999). 
\bibitem{KriLoi} A.~Krimmel, A.~Loidl, M.~Klemm, S.~Horn, and H.~Schrober, 
Phys. Rev. Lett. {\bf 82}, 2919 (1999). 
\bibitem{FYZG} P.~Fulde, A.N.~Yaresko, A.A.~Zvyagin, and Y.~Grin, Europhys. 
Lett. {\bf 54}, 779 (2001); 
\bibitem{LeeLV} J.D.~Lee, Phys. Rev. B {\bf 67}, 153108 (2003).  
\bibitem{YamU} Y.~Yamashita and K.~Ueda, Phys. Rev. B {\bf 67}, 195107 (2003); 
\bibitem{MatsUU} Y.~Matsushita, H.~Ueda, Y.~Ueda, Nature Materials {\bf 4}, 
845 (2005). 
\bibitem{ShiTak} Y.~Shimizu, H.~Takeda, M.~Tanaka, M.~Itoh, S.~Niitaka, and 
H.~Takagi, Nature Communications {\bf 3}, 981 (2012). 
\bibitem{mag} T.~Yamada, K.~Suzuki, and S.~Chikazumi, Appl. Phys. Lett. 
{\bf 13}, 172 (1968). 
\bibitem{Sam} E.J.~Samuelsen, E.J.~Bleeker, L.~Dobrzynski and T.~Riste, J. 
Appl. Phys. {\bf 39}, 1114 (1968). 
\bibitem{Chib} K.~Chiba, K.~Suzuki, and S.~Chikazumi, J. Phys. Soc. Jpn. Lett. 
{\bf 39}, 839 (1975). 
\bibitem{MiyShi} Y.~Miyamoto and M.~Shindo, J. Phys. Soc. Jpn. {\bf 62}, 1423 
(1993). 
\bibitem{WrAttRad} J.P.~Wright, J.P.~Attfield, and P.G.~Radaelli, Phys. Rev. 
Lett. {\bf 87}, 266401 (2001). 
\bibitem{revspin} P.~Fulde, J. Phys.: Condens. Matt. {\bf 16}, S591 (2003). 
\bibitem{Lac} C.~Lacroix, Physica B {\bf 404}, 3038 (2009). 
\bibitem{LeeTak} S.H.~Lee, H.~Takagi, D.~Louca, M.~Matsuda, S.~Ji, H.~Ueda, 
Y.~Ueda, T.~Katsufuji, J.H.~Chung, S.~Park, S.W.~Cheong, and C.~Broholm, 
J. Phys. Soc. Jpn. {\bf 79}, 011004 (2010). 
\bibitem{TakU} H.~Takagi and S.~Niitaka, {\em Highly Frustrated Magnetism in 
Spinels}, In: {\em Introduction to frustrated magnetism: Materials, 
experiments, theory}, Eds. C.~Lacroix, P.~Mendels, F.~Mila, Springer Series in 
Solid-State Sciences, {\bf 164}, 155 (2011).
\bibitem{Vil} J.~Villain, Z. Phys. B {\bf 33}, 31 (1979). 
\bibitem{Sub} M.A.~Subramanian, G.~Aravamudan and G.V.~Subba Rao, Progr. Sol. 
State Chem. {\bf 15}, 55 (1983). 
\bibitem{Gree} J.E.~Greedan in: {\em Magnetic Properties of Nonmetallic 
Compounds Based on Transition Elements}, Ed. H.P.J.~Wijn, Ladolt-B\"ornstein, 
New Series, {\bf 27}, 100 Springer Verlag, Berlin (1992).  
\bibitem{G} J.E.~Greedan, J. Alloys Compd. {\bf 408-412} 444 (2006).
\bibitem{GGG} J.S.~Gardner, M.J.P.~Gingras, and J.E.~Greedan, Rev. Mod. Phys. 
{\bf 82}, 53 (2010). 
\bibitem{Ginrev} M.J.P.~Gingras, {\em Spin Ice}, in {\em Highly Frustrated 
Magnetism}, Eds.: C.~Lacroix, P.~Mendels, and F.~Mila, Springer, Berlin (2010).
\bibitem{single} S.~Rosenkranz, A.P.~Ramires, A.~Hayashi, R.J.~Cava, 
R.~Siddharthan, and B.S.~Shastry, J. Appl. Phys. {\bf 87}, 5914 (2000). 
\bibitem{GinHerFau} M.J.P.~Gingras, B.C.~den Hertog, M.~Faucher, J.S.~Gardner, 
S.R.~Dunsiger, L.J.~Chang, B.D.~Gaulin, N.P.~Raju, and J.E.~Greedan, Phys. 
Rev. B {\bf 62}, 6496 (2000). 
\bibitem{YanGho} Y.M.~Yana and D.~Ghosh, Phys. Rev. B {\bf 61}, 9657 (2000). 
\bibitem{YanSenGho} Y.M.~Jana, A.~Sengupta, and D.~Ghosh, J. Magn. Magn. Mater. 
{\bf 248}, 7 (2002). 
\bibitem{MirBonHen} I.~Mirebeau, P.~Bonville, and M.~Hennion, Phys. Rev. B 
{\bf 76}, 184436 (2007). 
\bibitem{Hutch} M.T.~Hutchings, Solid State Phys. {\bf 16}, 227 (1964).   
\bibitem{Ryab} I.D.~Ryabov, J. Magn. Reson. {\bf 140}, 141 (1999). 
\bibitem{RudChu} C.~Rudowicz and C.Y.~Chung, J. Phys.: Condens. Matter 
{\bf 16}, 5825 (2004).
\bibitem{exch} S.H.~Curnoe, Phys. Rev. B {\bf 78}, 094418 (2008). 
\bibitem{McCCurGin} P.A.~McClarty, S.H.~Curnoe, and M.J.P.Gingras, J.Phys.: 
Conf. Ser. {\bf 145}, 012032 (2009). 
\bibitem{HG}  B.C.~den Hertog and M.J.P.~Gingras, Phys. Rev. Lett. {\bf 84}, 
3430 (2000).
\bibitem{exch1} T.~Yavorskii, T.~Fennell, M.P.J.~Gingras, and S.T.~Bramwell, 
Phys. Rev. Lett. {\bf 101}, 037204 (2008). 
\bibitem{M98} R.~Moessner, Phys. Rev. B {\bf 57}, R5587 (1998). 
\bibitem{MC} R.~Moessner and J.T.~Chalker, Phys. Rev. B {\bf 58}, 12049 (1998). 
\bibitem{Sha} R.~Siddharthan, B.S.~Shastry, and A.P.~Ramirez, Phys. Rev. B 
{\bf 63}, 184412 (2001). 
\bibitem{B1} S.T.~Bramwell, M.J.~Harris, B.C.~den Hertog, M.J.P.~Gingras, 
J.S.~Gardner, D.F.~McMorrow, A.R.~Wildes, A.L.~Cornelius, J.D.M.~Champion, 
R.G.~Melko, and T.~Fennell, Phys. Rev. Lett. {\bf 87}, 047205 (2001). 
\bibitem{Sid} R.~Siddharthan, B.S.~Shastry, A.P.~Ramirez, A.~Hayashi, 
R.J.~Cava, and S.~Rosenkranz, Phys. Rev. Lett. {\bf 83}, 1854 (1999). 
\bibitem{Ew} C.~Kittel, {\em Introduction to Solid State Physics}, Willey, NY 
(2005).
\bibitem{MG} R.G.~Melko and M.J.P.~Gingras, J. Phys.: Condens. Matter {\bf 16}, 
R1277 (2004);  
\bibitem{BarNew} G.T.~Barkema, and M.E.J.~Newman, Phys. Rev. E {\bf 57}, 1155 
(1998). 
\bibitem{MHG} R.G.~Melko, B.C.~den Hertog, and M.J.P.~Gingras, Phys. Rev. Lett. 
{\bf 87}, 067203 (2001). 
\bibitem{RufMelGin} J.P.C.~Ruff, R.G.~Melko, and M.G.P.~Gingras, Phys. 
Rev. Lett. {\bf 95}, 097202 (2005). 
\bibitem{mf} J.N.~Reimers, A.J.~Berlinsky, and A.C.~Shi, Phys. Rev. B 
{\bf 43}, 865 (1991). 
\bibitem{GinHer1} M.J.P.~Gingras and B.C.~den Hertog, Can.J. Phys. 
{\bf 79}, 1339 (2001). 
\bibitem{IsMoSon} S.V.~Isakov, R.~Moessner, and S.L.~Sondhi, Phys. Rev. 
Lett. {\bf 95}, 217201 (2005). 
\bibitem{Fuz} H.~Fukazawa, R.G.~Melko, R.~Higashinaka, Y.~Maeno, and 
M.J.P.~Gingras, Phys. Rev. B {\bf 65}, 054410 (2002). 
\bibitem{Is1} S.V.~Isakov, K.~Gregor, R.~Moessner, and S.L.~Sondhi, Phys. Rev. 
Lett. {\bf 93}, 167204 (2004). 
\bibitem{Coul} C.L.~Henley, Phys. Rev. B {\bf 71}, 014424 (2005). 
\bibitem{KheMoParSon} V.~Khemani, R.~Moessner, S.A.~Parameswaran, and 
S.L.~Sondhi, Phys. Rev. B {\bf 86}, 054411 (2012). 
\bibitem{FDWSPBAMB} T.~Fennell, P.P.~Deen, A.R.~Wildes, K.~Schmalzl, 
D.~Prabhakaran, A.T.~Boothroyd, R.J.~Aldus, D.F.~McMorrow, and S.T.~Bramwell, 
Science {\bf 326}, 415 (2009). 
\bibitem{H1} M.J.~Harris, S.T.~Bramwell, D.F.~McMorrow, T.~Zeiske, and 
K.W.~Godfrey, Phys. Rev. Lett. {\bf 79}, 2554 (1997). 
\bibitem{Har} M.J.~Harris, S.T.~Bramwell, T.~Zeiske, D.F.~McMorrow, and 
P.J.C.~King, J. Magn. Magn. Mater. {\bf 177}, 757 (1998). 
\bibitem{BraGing} S.T.~Bramwell and M.J.P.~Gingras, Science {\bf 294}, 
1495  (2001). 
\bibitem{R1} A.P.~Ramirez, A.~Hayashi, R.J.~Cava, R.~Siddharthan, and 
B.S.~Shastry, Nature {\bf 399}, 333 (1999).  
\bibitem{Kan} M.~Kanada, Y.~Yasui, Y.~Kondo, S.~Iikubo, M.~Ito, H.~Harashina, 
M.~Sato, H.~Okumura, K.~Kakurai, and H.~Kadowaki, J. Phys. Soc. Jpn. {\bf 71}, 
313 (2002).
\bibitem{Kad} H.~Kadowaki, Y.~Ishii, K.~Matsuhira, and Y.~Hinatsu, 
Phys. Rev. B {\bf 65}, 144421 (2002).
\bibitem{Pet1} T.~Fennell, O.A.~Petrenko, B.~Fak, J.S.~Gardner, S.T.~Bramwell, 
and B.~Ouladdiaf, Phys. Rev. B {\bf 72}, 224411 (2005). 
\bibitem{BY} K.~Binder and A.P.~Young, Rev. Mod. Phys. {\bf 58}, 801 (1986).
\bibitem{BH} S.T.~Bramwell and M.J.~Harris, J. Phys.: Condens. Matter {\bf 10}, 
L215 (1998).
\bibitem{Higa} R.~Higashinaka, H.~Fukuzawa, D.~Yanagishima, and Y.~Maeno, J. 
Phys. Chem Solids {\bf 63}, 1043 (2002). 
\bibitem{Hir} Z.~Hiroi, K.~Matsuhira, S.~Takagi, T.~Tayama, and T.~Sakakibara, 
J. Phys. Soc. Jpn. {\bf 72}, 411 (2003). 
\bibitem{Ke} X.~Ke, R.S.~Freitas, B.G.~Ueland, G.C.~Lau, M.L.~Dahlberg, 
R.J.~Cava, R.~Moessner, and P.~Schiffer, Phys. Rev. Lett. {\bf 99}, 137203 
(2007).   
\bibitem{CG} A.L.~Cornelius and J.S.~Gardner, Phys. Rev. B {\bf 64}, 060406 
(2001). 
\bibitem{Ke1} X.~Ke, B.G.~Ueland, D.V.~West, M.L.~Dahlberg, R.J.~Cava, and 
P.~Schiffer, Phys. Rev. B {\bf 76}, 214413 (2007). 
\bibitem{Mats} K.~Matsuhira, Y.~Hinatsu, K.~Tenya, and T.~Sakakibara, J. Phys.: 
Condens. Matter {\bf 12}, L649 (2000). 
\bibitem{Mats1} K.~Matsuhira, C.~Sekine, C.~Paulsen, and Y.~Hinatsu, J. Magn. 
Magn. Mater. {\bf 272-276}, E981 (2004). 
\bibitem{Blo} H.W.J.~Bl\"ote, R.F.~Wielinga, and W.J.~Huiskamp, Physica 
{\bf 43}, 549 (1969). 
\bibitem{Flood} D.J.~Flood, J. Appl. Phys. {\bf 45}, 4041 (1974). 
\bibitem{Tim} P.N.~Timonin, J. Exp. Theor. Phys. {\bf 113}, 251 (2011). 
\bibitem{HMTTS} Z.~Hiroi, K.~Matsuhira, S.~Takagi, T.~Tayama, and 
T.~Sakakibara, J. Phys. Soc. Jpn. {\bf 72}, 411 (2003). 
\bibitem{HFM} R.~Higashinaka, H.~Fukazawa, and Y.~Maeno, Phys. Rev. B 
{\bf 68}, 014415 (2003).  
\bibitem{IRMS} S.V.~Isakov, K.S.~Raman, R.~Moessner, and S.L.~Sondhi, Phys. 
Rev. B {\bf 70}, 104418 (2004).
\bibitem{MoSon} R.~Moessner and S.L.~Sondhi, Phys. Rev. Lett. {\bf 68}, 064411 
(2003).
\bibitem{Kast1} L.~Jaubert, J.T.~Chalker, P.C.W.~Holdsworth, and R.~Moessner, 
Phys. Rev. Lett. {\bf 100}, 067207 (2008). 
\bibitem{Kast} P.W.~Kasteleyn, Physica {\bf 27}, 1209 (1961). 
\bibitem{Kasta} P.W.~Kasteleyn, J. Math. Phys. {\bf 4}, 287 (1963).  
\bibitem{FBMMW} T.~Fenell, S.T.~Bramwell, D.F.~McMorrow, P.~Manuel, 
and A.R.~Wildes, Nature Physics {\bf 3}, 566 (2007).
\bibitem{HMO} Z.~Hiroi, K.~Matsuhira, and M.~Ogata, J.Phys. Soc. Jpn. 
{\bf 72}, 3045 (2003). 
\bibitem{YosNeWa} S.~Yoshida, K.~Nemoto, and K.~Wada, J. Phys. Soc. Jpn. 
{\bf 73}, 1619 (2004).  
\bibitem{magn} K.~Matsuhira, Z.~Hiroi, T.~Tayama, S.~Takagi, and 
S.~Sakakibara, J. Phys.: Condens. Matter {\bf 14}, L559 (2002). 
\bibitem{PetLeeBal} O.A.~Petrenko, M.R.~Lees, and G.~Balakrishnan, Phys. Rev. 
B {\bf 68}, 012406 (2003). 
\bibitem{Sak} T.~Sakakibara, T.~Tayama, Z.~Hiroi, K.~Matsuhira, and S.~Takagi, 
Phys. Rev. Lett. {\bf 90}, 207205 (2003).  
\bibitem{HiFuMa} R.~Higashinaka, H.~Fukazawa, and Y.~Maeno, Physica B 
{\bf 329-333}, 1040 (2003). 
\bibitem{Aok} H.~Aoki, T.~Sakakibara, K.~Matsuhira, and Z.~Hiroi, J. Phys. 
Soc. Jpn. {\bf 73}, 2851 (2004). 
\bibitem{Hig} R.~Higashinaka, H.~Fukazawa, K.~Deguchi, and Y.~Maeno, J.Phys. 
Soc. Jpn. {\bf 73}, 2845 (2004). 
\bibitem{SaiHigMa} M.~Saito, R.~Higashinaka, and Y.~Maeno, Phys. Rev. B 
{\bf 72}, 144422 (2005). 
\bibitem{HigMa} R.~Higashinaka and Y.~Maeno, Phys. Rev. Lett. {\bf 95}, 237208 
(2005). 
\bibitem{Tab} Y.~Tabata, H.~Kadowaki, K.~Matsuhira, Z.~Hiroi, N.~Aso, 
E.~Ressouche, and B.~Fak, Phys. Rev. Lett. {\bf 97}, 257205 (2006). 
\bibitem{Clan} J.P.~Clancy, J.P.C.~Ruff, S.R.~Dunsiger, Z.~Zhao, 
H.A.~Dabkowska, J.S.~Gardner, Y.~Qui, J.R.D.~Copley, T.~Jenkins, and 
B.D.~Gaulin, Phys. Rev. B {\bf 79}, 014408 (2009). 
\bibitem{Mal} B.Z.~Malkin, T.T.A.~Lummen, P.H.M.~van Loosdrecht, G.~Dhalenne, 
and A.R.~Zakirov, J. Phys.: Condens. Matter {\bf 22}, 276003 (2010). 
\bibitem{PetLeeBal1} O.A.~Petrenko, M.R.~Lees, and G.~Balakrishnan, J. Phys.: 
Condens. Matter {\bf 23}, 164218 (2011). 
\bibitem{Krey} C.~Krey, S.~Legl, S.R.~Dunsiger, M.~Meven, J.S.~Gardner, 
J.M.~Roper, and C.~Pfleiderer, Phys. Rev. Lett. {\bf 108}, 257204  (2012). 
\bibitem{Matt} M.J.~Matthews, C.~Castelnovo, R.~Moessner, S.A.~Grigera, 
D.~Prabhakaran, and P.~Schiffer, Phys. Rev. B {\bf 86}, 214419 (2012). 
\bibitem{Har1} M.J.~Harris, S.T.~Bramwell, P.C.W.~Holdsworth, and J.D.~Champion,
Phys. Rev. Lett. {\bf 81}, 4496 (1998). 
\bibitem{Fuk} H.~Fukazawa and Y.~Maeno, J. Phys. Soc. Jpn. {\bf 71}, 2578 
(2002). 
\bibitem{Mats2} K.~Matsuhira, Y.~Hinatsu, and T.~Sakakibara, J. Phys.: 
Condens. Matter {\bf 13}, L737 (2001). 
\bibitem{Sny} J.~Snyder, J.L.~Slusky,R.L.~Cava, and P.~Schiffer, Nature 
{\bf 413}, 48 (2001). 
\bibitem{Sny1} J.~Snyder, B.G.~Ueland, J.S.~Slusky, H.~Karunadasa, R.J.~Cava, 
A.~Mizel, and P.~Schiffer, Phys. Rev. Lett. {\bf 91}, 107201 (2003). 
\bibitem{Sny2} J.~Snyder, B.G.~Ueland, J.S.~Slusky, H.~Karunadasa, 
R.J.~Cava, A.~Mizel, and P.~Schiffer, Phys. Rev. B {\bf 69}, 064414 (2004). 
\bibitem{Sny3} J.~Snyder, B.G.~Ueland, A.~Mizel, J.S.~Slusky, H.~Karunadasa, 
R.J.~Cava, and P.~Schiffer, Phys. Rev. B {\bf 70}, 184431 (2004). 
\bibitem{Uel} B.G.~Ueland, G.C.~Lau, R.J.~Cava, J.R.~O'Brien, and P.~Schiffer, 
Phys. Rev. Lett. {\bf 96}, 027216 (2006); 
\bibitem{Sut} J.P.~Sutter, S.~Tsutsui, R.~Higashiraka, Y.~Maeno, O.~Leupold, 
and A.Q.R.~Baron, Phys. Rev. B {\bf 75}, 140402 (2007). 
\bibitem{Kit} K.~Kitagawa, R.~Higashinaka, K.~Ishida, Y.~Maeno, and 
M.~Takigawa, Phys. Rev. B {\bf 77}, 214403 (2008). 
\bibitem{Slob} D.~Slobinsky, C.~Castelnovo, R.A.~Borzi, A.S.~Gibbs, 
A.P.~Mackenzie, R.~Moessner, and S.A.~Grigera, Phys. Rev. Lett. {\bf 105}, 
267205 (2010). 
\bibitem{Quil} J.A.~Quilliam, L.R.~Yarashkavitch, H.A.~Dabkowska, B.D.~Gaulin, 
and J.B.~Kycia, Phys. Rev. B {\bf 83}, 094424 (2011). 
\bibitem{Klem} B.~Klemke, M.~Meissner, P.~Strehlow, K.~Kiefer, S.A.~Grigera, 
and D.A.~Tennant, J. Low Temp. Phys. {\bf 163}, 345 (2011). 
\bibitem{LinLiaChe} C.J.~Lin, C.N.~Liao, and C.H.~Chern, Phys. Rev. B 
{\bf 85}, 134434 (2012). 
\bibitem{musr} J.~Lago, S.J.~Blundell, and C.~Baines, J. Phys.: Condens. Matter 
{\bf 19}, 326210 (2007). 
\bibitem{Que} P.~Quemerais, P.~McClarty, and R.~Moessnner, Phys. Rev. Lett. 
{\bf 109}, 127601 (2012).
\bibitem{Ehl} G.~Ehlers, A.L.~Cornelius, M.~Orendac, M.~Kajnakova, T.~Fennell, 
S.T.~Bramwell, and J.S.~Gardner, J. Phys.: Condens. Matter {\bf 15}, L9 (2003). 
\bibitem{Erf} S.~Erfanifam, S.~Zherlitsyn, J.~Wosnitza, R.~Moessner, 
O.A.~Petrenko, G.~Balakrishnan, and A.A.~Zvyagin, Phys. Rev. B {\bf 84}, 
220404(R) (2011).
\bibitem{calor} M.~Orendac, J.~Hanko, E.~Cizmar, A.~Orendacova, M.~Shirai, and 
S.T.~Bramwell, Phys. Rev. B {\bf 75}, 104425 (2007).
\bibitem{nucl} G.~Ehlers, E.~Mamontov, M.~Zamponi, K.C.~Kam, and J.S.~Gardner, 
Phys. Rev. Lett. {\bf 102}, 016405 (2009). 
\bibitem{elast} Y.~Nakanishi, T.~Kumagai, M.~Yoshizawa, K.~Matsuhira, 
S.~Takagi, and Z.~Hiroi, Phys. Rev. B {\bf 83}, 184434 (2011). 
\bibitem{Sn} T.J.~Snee, R.E.~Meads, and W.G.~Parker, J. Phys. C {\bf 10}, 1761 
(1977); K.~Matsuhira, Y.~Hinatsu, K.~Tenya, H.~Amitsuka, and T.~Sakakibara, 
J. Phys. Soc. Jpn. {\bf 71}, 1576 (2002). 
\bibitem{Ehl1} G.~Ehlers, A.~Huq, S.O.~Diallo, C.~Adriano, K.C.~Rule, 
A.L.~Cornelius, P.~Fouquet, P.G.~Pagliuso, and J.S.~Gardner, J. Phys.: 
Condens. Matt. {\bf 24}, 076005 (2012). 
\bibitem{Zhou} H.D.~Zhou, J.G.~Cheng, A.M.~Hallas, C.R.~Wiebe, G.~Li, 
L.~Balicas, J.S.~Zhou, J.B.~Goodenough, J.S.~Gardner, and E.S.~Choi, Phys. 
Rev. Lett. {\bf 108}, 207206 (2012). 
\bibitem{CMS} C.~Castelnovo, R.~Moessner, and S.L.~Sondhi, Nature {\bf 451}, 
42 (2008). 
\bibitem{Lee1} S.-H.~Lee, C.~Broholm, W.~Ratcliff, G.~Gasparovic, Q.~Huang, 
T.H.~Kim, and S.-W.~Cheong, Nature {\bf 418}, 856 (2002). 
\bibitem{Dir} P.A.M.~Dirac, Proc. Roy. Soc. A {\bf 133}, 60 (1931). 
\bibitem{AB} Y.~Aharonov and D.~Bohm, Phys. Rev. {\bf 115}, 485 (1959). 
\bibitem{AC} Y.~Aharonov and A.~Casher, Phys. Rev. Lett. {\bf 53}, 319 (1984). 
\bibitem{Mil} K.A.~Milton, Rep. Prog. Phys. {\bf 69}, 1637 (2006). 
\bibitem{Ryzh} I.A.~Ryzhkin, J. Exp. Theor. Phys. {\bf 101}, 481 (2005).  
\bibitem{Nag2} J.F.~Nagle, Chem. Phys.{\bf 43}, 317 (1979). 
\bibitem{Jack} J.D.~Jackson, {\em Classical Electrodynamics}, Wiley, NY (1975). 
\bibitem{Fish} V.~Kobelev, A.B.~Kolomeisky and M.E.~Fisher, J. Chem. Phys. 
{\bf 116}, 7589 (2002). 
\bibitem{Li2} O.J.~Heilmann and E.H.~Lieb, Comm. Math. Phys. {\bf 25}, 190 
(1972). 
\bibitem{Fang} Z.~Fang, N.~Nagaosa, K.S.~Takahashi, A.~Asamitsu, R.~Mathieu,
T.~Ogasawara, H.~Yamada, M.~Kawasaki, Y.~Tokura, and K.~Terakura, Science 
{\bf 302}, 92 (2003). 
\bibitem{Bram1} S.T.~Bramwell, Phil. Trans. Roy. Soc. A {\bf 370}, 5738 (2012).
\bibitem{MoShif} R.~Moessner and P.~Schiffer, Nature Physics {\bf 5}, 250 
(2009). 
\bibitem{JH} L.D.C.~Jaubert and P.C.W.~Holdsworth, Nature Physics {\bf 5}, 
258 (2009).
\bibitem{JH1} L.D.C.~Jaubert and P.C.W.~Holdsworth, J. Phys.: Condens. Matter 
{\bf 23}, 164222 (2011).
\bibitem{Ons} L.~Onsager, J. Chem. Phys. {\bf 2}, 599 (1934). 
\bibitem{Wien} S.T.~Bramwell, S.R.~Giblin, S.~Calder, R.~Aldus, D.~Prabhakaran, 
and T.~Fennell, Nature {\bf 461}, 956 (2009).
\bibitem{Gibl} S.R.~Giblin, S.T.~Bramwell, P.C.W.~Holdsworth, D.~Prabhakaran, 
and I.~Terry, Nature Physics {\bf 7}, 252 (2011). 
\bibitem{Dun} S.R.~Dunsiger, A.A.~Aczel, C.~Arguello, H.A.~Dabkowska, 
A.~Dabkowski, M.H.~Du, T.~Goko, B.~Javanparast, T.~Lin, F.L.~Ning, 
H.M.L.~Noad, D.J.~Singh, T.J.~Williams, Y.~J.~Uemura, M.J.P.~Gingras, and 
G.M.~Luke, Phys. Rev. Lett. {\bf 107}, 207207 (2011). 
\bibitem{Bram} S.T.~Bramwell, J. Phys.: Condens. Matter {\bf 23}, 112201 
(2011).
\bibitem{StyFei} A.V.~Shtyk and M.V.~Feigelman, JETP lett. {\bf 92}, 799 
(2010).
\bibitem{RyzRyz} I.A.~Ryzhkin and M.I.~Ryzhkin, JETP Lett. {\bf 93}, 384 
(2011). 
\bibitem{RyzKlu} I.A.~Ryzhkin, A.V.~Klyuev, M.I.~Ryzhkin, and I.V.~Tsybulin, 
JETP Lett. {\bf 95}, 302 (2012).
\bibitem{Blu} S.J.~Blundell, Phys. Rev. Lett. {\bf 108}, 147601 (2012). 
\bibitem{TakTat} A.~Takeuchi and G.~Tatara, J. Appl. Phys. {\bf 111}, 07C509 
(2012). 
\bibitem{Stre} P.~Strehlow, S.~Neubert, B.~Klemke, and M.~Meissner, Cont. 
Mech. Therm. {\bf 24}, 347 (2012).   
\bibitem{Gin} M.J.P.~Gingras, Science {\bf 326}, 375 (2009).
\bibitem{Morr} D.J.T.~Morris, D.A.~Tennant, S.A.~Grigera, B.~Klemke, 
C.~Castelnovo, R.~Moessner, C.~Czternasty, M.~Meissner, K.C.~Rule, 
J.-U.~Hoffmann, K.~Kiefer, S.~Gerischer, D.~Slobinsky, and R.S.~Perry, 
Science {\bf 326} 411 (2009). 
\bibitem{FenDee} T.~Fennell, P.P.~Deen, A.R.~Wildes, K.~Schmalzl, 
D.~Parabharkan, A.T.~Boothroyd, R.J.~Aldus, D.F.~McMarrow, and S.T.~Bramwell, 
Science, {\bf 326} 415 (2009). 
\bibitem{Kadow} H.~Kadowaki, N.~Doi, Y.~Aoki, Y.~Tabata, T.J.~Sato, J.W.~Lynn, 
K.~Matsuhira, and Z.~Hiroi, J. Phys. Soc. Jpn. {\bf 78}, 103706 (2009).
\bibitem{CMS3} C.~Castelnovo, R.~Moessner, and S.L.~Sondhi, Phys. Rev. Lett. 
{\bf 104}, 107201 (2010).
\bibitem{Zhou1} H.D.~Zhou, S.T.~Bramwell, J.G.~Cheng, C.R.~Wiebe, 
G.~Li, L.~Balicas, J.A.~Bloxsom, H.J.~Silverstein, J.S.~Zhou, J.B.~Goodenough, 
and J.S.~Gardner, Nature Commun. {\bf 2}, 478 (2011). 
\bibitem{CMS2} C.~Castelnovo, R.~Moessner, and S.L.~Sondhi, Phys. Rev. B {\bf 
84}, 144435 (2011). 
\bibitem{other} J.A.~Quilliam, L.R.~Yaraskavitch, H.A.~Dabkowska, B.D.~Gaulin, 
and J.B.~Kycia, Phys. Rev. B {\bf 83}, 094424 (2011). 
\bibitem{Yar} L.R.~Yaraskavitch, H.M.~Revell, S.~Meng, K.A.~Ross, H.M.L.~Noad, 
H.A.~Dabkowska, B.D.~Gaulin, and J.B.~Kycia, Phys. Rev. B {\bf 85}, 020410(R) 
(2012). 
\bibitem{Rev} H.M.~Revell, L.R.~Yaraskavitch, J.D.~Mason, K.A.~Ross, 
H.M.L.~Noad, H.A.~Dabkowska, B.D.~Gaulin, P.~Henelius, and J.B.~Kycia, 
Nature Physics {\bf 9}, 34 (2013).  
\bibitem{Sala} G.~Sala, C.~Castelnovo, R.~Moessner, S.L.~Sondhi, K.~Kitagawa, 
M.~Takigawa, R.~Hagashinaka, and Y.~Maeno, Phys. Rev. Lett. {\bf 108}, 217203 
(2012).
\bibitem{Koll} G.~Kolland, O.~Breunig, M.~Valldor, M.~Hiertz, 
J.~Frielingsdorf, T.~Lorenz, Phys. Rev. B {\bf 86}, 060402(R) (2012). 
\bibitem{Moe} R.~Moessner, Can. J. Phys. {\bf 79}, 1283 (2001). 
\bibitem{qsi} A.G.~Del Maestro and M.J.P.~Gingras, J. Phys.: Condens. Matter 
{\bf 16}, 3339 (2004). 
\bibitem{MoeTcheSon} R.~Moessner, O.~Tchernyshyov, and S.L.~Sonhi, J. Stat. 
Phys. {\bf 116}, 755 (2004). 
\bibitem{Bur} F.J.~Burnell, S.~Chakravarty, and S.L.~Sondhi, Phys. Rev. B 
{\bf 79}, 144432 (2009). 
\bibitem{MaeOySa} S.~Maegawa, A.~Oyamada, and S.~Sato, J. Phys. Soc. Jpn. 
{\bf 79}, 011002 (2010). 
\bibitem{OnTa} S.~Onoda and Y.~Tanaka, Phys. Rev. B {\bf 83}, 094411 (2011). 
\bibitem{Ross} K.A.~Ross, L.~Savary, B.D.~Gaulin, and L.~Balents, 
Phys. Rev. X {\bf 1}, 021002 (2011). 
\bibitem{Shan} N.~Shannon, O.~Sikora, F.~Pollmann, K.~Penc, and P.~Fulde, 
Phys. Rev. Lett. {\bf 108}, 067204 (2012). 
\bibitem{WanTch} Y.~Wan and O.~Tchernyshyov, Phys. Rev. Lett. {\bf 108}, 247210 
(2012). 
\bibitem{BenSikSha} O.~Benton, O.~Sikora, and N.~Shannon, Phys. Rev. B 
{\bf 86}, 075154 (2012). 
\bibitem{LeeOnBal} S.~Lee, S.~Onoda, and L.~Balents, Phys. Rev. B {\bf 86}, 
104412 (2012).
\bibitem{Yb} J.A.~Hodges, P.~Bonville, A.~Forget, M.~Rams, K.~Krolas, and 
G.~Dhalenne, J. Phys. Condens. Matter {\bf 13}, 9301 (2001). 
\bibitem{Hod} J.A.~Hodges, P.~Bonville, A.~Forget, A.~Yaouanc, P.~Dalmas de 
Reotier, G.~Andre, M.~Rams, K.~Krolas, C.~Ritter, P.C.M.~Gubbens, C.T.~Kaiser, 
P.J.C.~King, and C.~Baines, Phys. Rev. Lett. {\bf 88}, 077204 (2002). 
\bibitem{Ross1} K.A.~Ross, J.P.C.~Ruff, C.P.~Adams, J.S.~Gardner, 
H.A.~Dabkowska, Y.~Qiu, J.R.D.~Copley, and B.D.~Gaulin, Phys. Rev. Lett. 
{\bf 103}, 227202 (2009). 
\bibitem{Yas} Y.~Yasui, M.~Soda, S.~Iikubo, M.~Ito, M.~Sato, N.~Hamaguchi, 
T.~Matsushita, N.~Wada, T.~Takeuchi, N.~Aso, and K.~Kakurai, J. Phys. Soc. 
Jpn. {\bf 72}, 3014 (2003). 
\bibitem{Gar} J.S.~Gardner, G.~Ehlers, N.~Rosov, R.W.~Erwin, and C.~Petrovic, 
Phys. Rev. B {\bf 70}, 180404(R) (2004). 
\bibitem{Cao} H.B.~Cao, A.~Gukasov, I.~Mirebeau, and P.~Bonville, J. Phys.: 
Condens. Matter {\bf 21}, 492202 (2009). 
\bibitem{Yao} A.~Yaouanc, P.~Dalmas de Reotier, C.~Marin, and V.~Glazkov, 
Phys. Rev. B {\bf 84}, 172408 (2011). 
\bibitem{Ross2} K.A.~Ross, L.R.~Yaraskavitch, M.~Laver, J.S.~Gardner, 
J.A.~Quilliam, S.~Meng, J.B.~Kycia, D.K.~Singh, Th.~Proffen, H.A.~Dabkowska, 
and B.D.~Gaulin, Phys. Rev. B {\bf 84}, 174442 (2011). 
\bibitem{Tom} J.D.~Thompson, P.A.~McClarty, H.M.~Ronnow, L.~P.~Regnault, 
A.~Sorge, and M.J.P.~Gingras, Phys. Rev. Lett. {\bf 106}, 187202 (2011). 
\bibitem{Tom1} J.D.~Thompson, P.A.~McClarty, and M.J.P.~Gingras, J. Phys.: 
Condens. Matter {\bf 23}, 164219 (2011). 
\bibitem{Ross3} K.A.~Ross, Th.~Proffen, H.A.~Dabkowska, J.A.~Quilliam, 
L.R.~Yaraskavitch, J.B.~Kycia, and B.D.~Gaulim, Phys. Rev. B {\bf 86}, 144424 
(2012). 
\bibitem{Lho} E.~Lhotel, C.~Paulsen, P.D.~de Reotier, A.~Yaouanc, C.~Marin, 
and S.~Vanishri, Phys. Rev. B {\bf 86}, 020410(R) (2012). 
\bibitem{Appl} R.~Applegate, N.R.~Hayre, R.R.P.~Singh, T.~Lin, A.G.R.~Day, and 
M.J.P.~Gingras, Phys. Rev. Lett. {\bf 109}, 097205 (2012).   
\bibitem{Pow} S.~Powell Phys. Rev. B {\bf 84}, 094437 (2011). 
\bibitem{Chang} L.J.~Chang, S.~Onoda, Y.X.~Su, Y.J.~Kao, K.D.~Tsuei, 
Y.K.~Yasui, K.~Kakurai, and M.R.~Lees, Nature Commun. {\bf 3}, 992 
(2012). 
\bibitem{Hostuf} G.C.~Lau, R.S.~Freitas, B.G.~Ueland, B.D.~Muegge, E.L.~Duncan, 
P.Schiffer, and R.J.~Cava, Nature Physics {\bf 2}, 249 (2006). 
\bibitem{Lau} G.C.~Lau, B.D.~Muegge, T.M.~McQueen, E.L.~Duncan, and R.J.~Cava, 
J. Solid State Chem. {\bf 179}, 3126 (2006). 
\bibitem{Lau1} G.C.~Lau, R.S.~Freitas, B.G.~Ueland, M.L.~Dahlberg, Q.~Huang, 
H.W.~Zandbergen, P.~Schiffer, and R.J.~Cava, Phys. Rev. B {\bf 76}, 054430 
(2007). 
\bibitem{Zhou2} H.D.~Zhou, C.R.~Wiebe, Y.J.~Jo, L.~Balicas, Y.~Qiu, 
J.R.D.~Copley, G.~Ehlers, P.~Fouquet, and J.S.~Gardner, J. Phys.: Condens. 
Matt. {\bf 10}, 342201 (2007).  
\bibitem{Lau2} G.C.~Lau, T.M.~McQueen, Q.~Huang, H.W.~Zandbergen, and 
R.J.~Cava, J. Solid State Chem. {\bf 181}, 45 (2008). 
\bibitem{Ehl2} G.~Ehlers, J.S.~Gardner, Y.~Qiu, P.~Fouquet, C.R.~Wiebe, 
L.~Balicas, and H.D.~Zhou, Phys. Rev. B {\bf 77}, 052404 (2008). 
\bibitem{Gar1} J.S.~Gardner, G.~Ehlers, P.~Fouquet, B.~Farago, and 
J.R.~Stewart, J. Phys.: Condens. Matter {\bf 23}, 164220 (2011). 
\bibitem{Ald} R.J.~Aldus, T.~Fennell, P.P.~Deen, E.~Ressouche, G.C.~Lau, 
R.J.Cava, and S.T.~Bramwell, New J. Phys. {\bf 15} 013022 (2013).   
\bibitem{Dystuf} B.G.~Ueland, G.C.~Lau, R.S.~Freitas, J.~Snyder, 
M.L.~Dahlberg, B.D.~Muegge, E.L.~Duncan, R.J.~Cava, and P.~Schiffer, 
Phys. Rev. B  {\bf 77}, 020405(R) (2008).  
\bibitem{met} D.~Yanagishima, and Y.~Maeno, J. Phys. Soc. Jpn. {\bf 70}, 2880 
(2001). 
\bibitem{Nak} S.~Nakatsuji, Y.~Machida, Y.~Maeno, T.~Tayama, T.~Sakakibara, 
J.~van Duijn, L.~Balicas, J.N.~Millican, R.T.~Macaluso, and J.Y.~Chan, Phys. 
Rev. Lett. {\bf 96}, 087204 (2006). 
\bibitem{Mach} Y.~Machida, S.~Nakatsuji, Y.~Maeno, T.~Tayama, T.~Sakakibara, 
and S.~Onoda, Phys. Rev. Lett. {\bf 98}, 057203 (2007).    
\bibitem{metteor} A.~Ikeda and H.~Kawamura, J. Phys. Soc. Jpn. {\bf 77} 073707 
(2008). 
\bibitem{UdaIshMot} M.~Udagawa, H.~Ishizuka, and Y.~Motome, Phys. Rev. Lett. 
{\bf 108}, 066406 (2012).  
\bibitem{IshUdMot} H.~Ishizuka, M.~Udagawa, and Y.~Motome, J. Phys. Soc. Jpn. 
{\bf 81}, 113706 (2012)
\bibitem{sH} Y.~Machida, S.~Nakatsuji, H.~Tonomura, T.~Tayama, T.~Sakakibara, 
J.~van Duijn, C.~Broholm, and Y.~Maeno, J. Phys. Chem. Solids {\bf 66}, 1435 
(2005). 
\bibitem{Mill} J.N.~Millican, R.T.~Macaluso, S.~Nakatsuji, Y.~Machida, 
Y.~Maeno, and J.Y.~Chan, Mater. Res. Bull. {\bf 42}, 928 (2007). 
\bibitem{Mach1} Y.~Machida, S.~Nakatsuji,S.~Onda, T.~Tayama, and 
T.~Sakakibara, Nature {\bf 463}, 210 (2010).  
\bibitem{Wa} R.F.~Wang, C.~Nisoli, R.S.~Freitas, J.~Li. W.~McConville, 
B.J.~Cooley, M.S.~Lund, N.~Samarth, C.~Leighton, V.H.~Crespi, and P.~Schiffer, 
Nature {\bf 439}, 303 (2006).
\bibitem{optic} A.~Libal, C.~Reichhardt, and C.J.O.~Reichhardt, 
Phys. Rev. Lett. {\bf 97}, 228302 (2006). 
\bibitem{Stru} J.~Struck, C.~\"Olschl\"ager, R.~Le Targat, P.~Soltan-Panahi, 
A.~Eckardt, M.~Lewenstein, P.~Windpassinger, and K.~Sengstock, Science 
{\bf 333}, 996 (2011). 
\bibitem{asi} C.~Nisoli, R.F.~Wang, J.~Li, W.F.~McConville, P.E.~Lammert, 
P.~Schiffer, and V.H.~Crespi, Phys. Rev. Lett. {\bf 98}, 217203 (2007). 
\bibitem{LePo} A.~Leon and J.~Pozo, J. Magn. Magn. Mater. {\bf 320}, 210 
(2008). 
\bibitem{QiBrCu} Y.~Qi, T.~Brintlinger, and J.~Cumings, Phys. Rev. B {\bf 77}, 
094418 (2008). 
\bibitem{Meng} E.~Mengotti, L.J.~Heyderman, A.~Fraile Rodriguez, A.~Bisig, 
L.~Le Guyader, F.~Nolting, and H.B.~Braun, Phys. Rev. B {\bf 78}, 144402 
(2008). 
\bibitem{Tra} A.~Trabesinger, Nature Physics {\bf 4}, 832 (2008). 
\bibitem{Lib} A.~Libal, C.J.O.~Reichhardt, and C.~Reichhardt, Phys. Rev. Lett. 
{\bf 102}, 237004 (2009). 
\bibitem{LiKe} J.~Li, X.~Ke, S.~Zhang, D.~Garand, C.~Nisoli, P.~Lammert, 
V.H.~Crespi, and P.~Schiffer, Phys. Rev. B {\bf 81}, 092406 (2010). 
\bibitem{BudPolSta} Z.~Budrikis, P.~Politi, and  R.L.~Stamps, Phys. Rev. Lett. 
{\bf 105}, 017201 (2010). 
\bibitem{Shu} A.~Schumann, B.~Sothmann, P.~Szary, and H.~Zabel, Appl. Phys. 
Lett. {\bf 97}, 022509 (2010).
\bibitem{Nic} C.~Nicoli, J.~Li, X.L.~Ke, D.~Garand, P.~Schiffer, and 
V.H.~Crespi, Phys. Rev. Lett. {\bf 105}, 047205 (2010). 
\bibitem{Tche} O.~Tchernyshyov, Nature Phys. {\bf 6}, 323 (2010). 
\bibitem{Mell} P.~Mellado, O.~Petrova, Y.C.~Shen, and O.~Tchernyshyov, Phys. 
Rev. Lett. {\bf 105}, 187206 (2010). 
\bibitem{LiWang} Y.~Li and T.X.~Wang, Phys. Lett. A {\bf 374}, 4475 (2010). 
\bibitem{LiZh} J.~Li, S.~Zhang, J.~Bartell, C.~Nisoli, X.~Ke, P.E.~Lammetr, 
V.H.~Crespi, and P.~Schiffer, Phys. Rev. B {\bf 82}, 134407 (2010). 
\bibitem{Rou} N.~Rouemaille, F.~Montaigne, B.~Canals, A.~Duluard, D.~Lacour, 
M.~Hehn, R.~Belkhou, O.~Fruchart, S.~el Moussaoui, A.~Bendounan, and 
F.~Maccherozzi, Phys. Rev. Lett. {\bf 106}, 057209 (2011). 
\bibitem{LiWa} Y.~Li, T.X.~Wang, H.Y.~Liu, X.F.~Dai, and G.D.~Liu, Phys. Lett. 
A {\bf 375}, 1548 (2011).
\bibitem{Dau} S.A.~Daunheimer, O.~Petrova, O.~Tchernyshyov, and J.~Cumings, 
Phys. Rev. Lett. {\bf 107}, 167201 (2011). 
\bibitem{Budr1} Z.~Budrikis, P.~Politi, and R.L.~Stamps, Phys. Rev. Lett. 
{\bf 107}, 217204 (2011). 
\bibitem{Rei} C.J.O.~Reichhardt, A.~Libal, and C.~Reichhardt, New J. Phys. 
{\bf 14}, 025006 (2012). 
\bibitem{Kap} V.~Kapaklis, U.B.~Arnalds, A.~Harman-Clarke, E.T.~Papaioannou, 
M.~Karimipour, P.~Korelis, A.~Taroni, P.C.W.~Holdsworth, S.T.~Bramwell, and 
B.~Hjorvarsson, New J. Phys. {\bf 14}, 035009 (2012). 
\bibitem{Scu} A.~Scumann, P.~Szari, E.Y.~Vedmedenko, and H.~Zabel, New J. 
Phys. {\bf 14}, 035015 (2012); 
\bibitem{Budr2} K.~Budrikis, K.L.~Livesey, J.P.~Morgan, J.~Akerman, A.~Stein, 
S.~Langridge, C.H.~Marrows, and R.L.~Stamps, New J. Phys. {\bf 14}, 035014 
(2012). 
\bibitem{Budr3} Z.~Budrikis, P.~Politi, and R.L.~Stamps, J. Appl. Phys. 
{\bf 111}, 07E109 (2012). 
\bibitem{Mol} L.A.S.~Mol, A.R.~Pereira, and W.A.~Moura-Melo, Phys. Rev. B 
{\bf 85}, 184410 (2012). 
\bibitem{ZhaLi} S.~Zhang, J.~Li, I.~Gilbert, J.~Bartell, M.J.~Erikson, Y.~Pan, 
P.E.~Lammert, C.~Nisoli, K.K.~Kohli, R.~Misra, V.H.~Crespi, N.~Samarth, 
C.~Leighton, and P.~Schiffer, Phys. Rev. Lett. {\bf 109}, 087201 (2012). 
\bibitem{Arn} U.B.~Arnalds, A.~Farhan, R.V.~Chopdekar, V.~Kapaklis, A.~Balan, 
E.T.~Papaioannou, M.~Ahlberg, F.~Nolting, L.J.~Heyderman, and B.~Hjorvarsson, 
Appl. Phys. Lett. {\bf 101}, 112404 (2012). 
\bibitem{Morg} J.P.~Morgan, J.~Akerman, A.~Stein, C.~Phatak, R.M.L.~Evans, 
S.~Langridge, and C.H.~Marrows, Phys. Rev. B {\bf 87}, 024405 (2013). 
\bibitem{dis} K.K.~Kohli, A.L.~Balk, J.~li, S.~Zhang, I.~Gilbert, P.E.~Lammert, 
V.H.~Crespi, P.~Schiffer, and N.~Samarth, Phys. Rev. B {\bf 84}, 180412(R) 
(2011).
\bibitem{Budr4} Z.~Budrkis, J.P.~Morgan, J.~Akerman, A.~Stein, P.~Politi, 
S.~Langridge, C.H.~Marrows, and R.L.~Stamps, Phys. Rev. Lett. {\bf 109}, 
037203 (2012). 
\bibitem{Lamm} P.E.~Lammert, X.L.~Ke, J.~Li, C.~Nisoli, D.M.~Garand, 
V.H.~Crespi, and P.~Schiffer, Nature Phys. {\bf 6}, 786 (2010). 
\bibitem{monart} G.~M\"oller and R.~Moessner, Phys. Rev. Lett. {\bf 96}, 
237202 (2006).  
\bibitem{Mol1} L.A.~Mol, R.L.~Silva, R.C.~Silva, A.R.~Pereira, 
W.A.~Moura-Melo, and B.V.~Costa, J. Appl. Phys. {\bf 106}, 063913 (2009). 
\bibitem{MolMo} G.~M\"oller and R.~Moessner, Phys. Rev. B {\bf 80}, 140409(R) 
(2009). 
\bibitem{RosFra} G.~Rosenberg and M.~Franz, Phys. Rev. B {\bf 82}, 035105 
(2010). 
\bibitem{Morg1} J.P.~Morgan, A.~Stein, S.~Langridge, and C.H.~Marrows, Nature 
Physics {\bf 7}, 75 (2011). 
\bibitem{Lad} S.~Ladak, D.~Read, T.~Tyliszczak, W.R.~Branford, and  
L.F.~Cohen, New J. Phys. {\bf 13}, 023023 (2011). 
\bibitem{Pha} C.~Phatak, A.K.~Petford-Long, O.~Heinonen, M.~Tanase, and 
M.~De Graef, Phys. Rev. B {\bf 83}, 174431 (2011). 
\bibitem{Lad1} S.~Ladak, D.E.~Read, W.R.~Branford, and L.F.~Cohen, New J. 
Phys. {\bf 13}, 063032 (2011). 
\bibitem{MorSteLan} J.P.~Morgan, A.~Stein, S.~Langridge, and C.H.~Marrows, 
New J. Phys. {\bf 13}, 105002 (2011). 
\bibitem{Sil} R.C.~Silva, F.S.~Nascimento, L.A.S.~Mol, W.A.~Moura-Melo,and 
A.R.~Pereira, New J. Phys. {\bf 14}, 015008 (2012); C.~Nisoli, New J. Phys. 
{\bf 14}, 035017 (2012). 
\bibitem{Hug} R.V.~Hugli, G,~Duff, B.~O'Conchuir, E.~Mengotti, L.J.~Heyderman, 
A.F.~Rodriguez, F.~Nolting, and H.B.~Braun, J. Appl. Phys. {\bf 111}, 07E103 
(2012). 
\bibitem{Pol} S.D.~Pollard, V.~Volkov, and Y.~Zhu, Phys. Rev. B {\bf 85}, 
180402(R) (2012); 
\bibitem{Sil1} R.C.~Silva, R.J.C.~Lopes, L.A.S.~Mol, W.A.~Moura-Melo, 
G.M.~Wysin, and A.R.~Pereira, Phys. Rev. B {\bf 87}, 014414 (2013). 
\bibitem{Meng1} S.~Ladak, D.E.~Reaf, G.K.~Perkins, L.F.~Cohen, and 
W.R.~Branford, Nature Phys. {\bf 6}, 359 (2010). 
\bibitem{Meng2} E.~Mengotti,L.J.~Heyderman, A.F.~Rodriguez, F.~Nolting, 
R.V.~Hugli, and H.B.~Braun, Nature Physics {\bf 7}, 68 (2011). 
\bibitem{Cum} J.~Cumings, Nature Physics {\bf 7}, 7 (2011). 
\bibitem{aval} J.P.~Setha, K.A.~Dahmen and C.R.~Myers, Nature {\bf 410}, 242 
(2001). 
\bibitem{artaHe} W.R.~Branford, S.~Ladak, D.E.~Read, K.~Zeissler, and 
L.F.~Cohen, Science, {\bf 335}, 1597 (2012). 
\end{thebibliography}
\end{document}